\documentclass[aps,pra,twocolumn,superscriptaddress,notitlepage,nofootinbib,longbibliography, colorlinks=true]{revtex4-2}
\usepackage{mathrsfs,graphicx,amsfonts,amssymb,amsmath,mathtools}
\usepackage[dvipsnames]{xcolor}
\usepackage[detect-all=true,group-minimum-digits=4]{siunitx}
\usepackage[colorlinks,allcolors= blue]{hyperref}
\usepackage{cleveref}
\newcommand{\tcb}[1]{\textcolor{blue}{#1}}

\begin{document}
	
	\title{Enhancing mechanical entanglement in molecular optomechanics}
	
	\author{E. Kongkui Berinyuy}
	\email{emale.kongkui@facsciences-uy1.cm}
	\affiliation{Department of Physics, Faculty of Science, University of Yaounde I, P.O.Box 812, Yaounde, Cameroon}
	
	\author{C. Tchodimou}
	\email{emale.kongkui@facsciences-uy1.cm}
	\affiliation{Department of Physics, Faculty of Science, University of Yaounde I, P.O.Box 812, Yaounde, Cameroon}
	
	\author{P. Djorwé}
	\affiliation{Department of Physics, Faculty of Science, 
		University of Ngaoundere, P.O.Box 454, Ngaoundere, Cameroon}
	\affiliation{Stellenbosch Institute for Advanced Study (STIAS), Wallenberg Research Centre at Stellenbosch University, Stellenbosch 7600, South Africa}
	
	\author{A.-H. Abdel-Aty}
	\affiliation{Department of Physics, College of Sciences, University of Bisha, Bisha 61922, Saudi Arabia}
	\affiliation{Physics Department, Faculty of Science, Al-Azhar University, Assiut 71524, Egypt}

	\author{K.S. Nisar}
	\affiliation{Department of Mathematics, College of Science and Humanities in Al-Kharj, Prince Sattam Bin Abdulaziz University, Al-Kharj 11942, Saudi Arabia}
	
	\author{S. G. Nana Engo}
	\email{serge.nana-engo@facsciences-uy1.cm}
	\affiliation{Department of Physics, Faculty of Science, University of Yaounde I, P.O.Box 812, Yaounde, Cameroon}
	
	\begin{abstract}
		We propose a scheme for enhancing bipartite quantum entanglement in a double-cavity molecular optomechanical system (McOM) incorporating an intracavity optical parametric amplifier (OPA). Utilizing a set of linearized quantum Langevin equations and numerical simulations, we investigate the impact of the OPA on both optical-vibration and vibration-vibration entanglement. Our key findings reveal a counterintuitive trade-off: while the OPA significantly enhances vibration-vibration entanglement, a critical resource for quantum memories and transducers, it simultaneously suppresses optical-vibration entanglement, essential for quantum state transfer and readout. We demonstrate that maximal vibration-vibration entanglement is achieved when the molecular collective vibrational modes are symmetrically populated, providing a clear experimental guideline for optimizing entanglement sources. In particular, the vibration-vibration entanglement generated in our OPA-enhanced McOM system exhibits remarkable robustness to thermal noise, persisting at temperatures approaching \SI{e3}{\kelvin}, significantly exceeding conventional optomechanical systems in terms of thermal robustness, and highlighting the potential for room temperature quantum information processing. These results establish a promising theoretical foundation for OPA-enhanced McOM systems as a robust and scalable platform for quantum technologies, paving the way for future experimental implementations and advanced quantum information processing applications.
	\end{abstract}

	\maketitle
	
	\section{Introduction} \label{sec:Intro}
	
	Quantum entanglement, the non-classical correlation between quantum systems, is not merely a fundamental curiosity but the cornerstone of transformative quantum technologies, underpinning advancements from quantum computing to secure communication. Realizing the full potential of these technologies hinges on our ability to precisely control and manipulate quantum entanglement, and cavity optomechanical (COM) systems, which masterfully intertwine light and mechanical motion, have emerged as a leading architecture for this endeavor. These systems have been intensively investigated, attracting significant theoretical~\cite{Lai2022,Tchodimou2017,Araya2023,Agasti2024,Ye2025,Chen2025} and experimental~\cite{Kotler2021,Clarke2020} interest, driven by the promise of revolutionizing quantum information processing~\cite{Rips2013,Blais2020}, quantum sensing~\cite{Xia2023,Brady2023,Djor2024}, and quantum metrology~\cite{Barzanjeh2021}. Within the broader landscape of COM systems, molecular cavity optomechanical (McOM) systems are emerging as a particularly compelling frontier. Capitalizing on molecular polaritons, McOM systems – particularly those employing nanocavities – offer a unique suite of advantages over their macroscopic counterparts: unprecedentedly strong light-matter interactions arising from the strong coupling of molecular vibrations to confined photons, access to high-frequency vibrational modes inherent to molecules, and consequently, enhanced optomechanical coupling rates~\cite{Roelli2016,esteban2022molecular,schmidt2017}. As highlighted in recent perspectives~\cite{Roelli2024}, McOM systems are uniquely positioned to bridge quantum optics and molecular spectroscopy, potentially enabling room-temperature operation and scalable quantum technologies~\cite{Roelli2024,Chikkaraddy2016,CarlonZambon2022}. There is a growing hope that McOM systems, leveraging these advantages, might be used to generate unprecedented levels of mechanical quantum entanglement, thereby boosting quantum technologies.
	
	Molecular COM systems uniquely bridge quantum optics and molecular spectroscopy, offering a fertile ground for engineering novel quantum states and hybrid quantum networks~\cite{Liu2019,Huang2024,Roelli2024,Xiang2024}. A particularly intriguing prospect within these systems is the generation and manipulation of vibration-vibration entanglement, a pivotal resource for quantum memory and transduction~\cite{Xiang2024}. Although entanglement in macroscopic COM systems has been extensively explored~\cite{EMALE2025416919,Massembele2024,Tchodimou2017,Djo2024}, the potential of nonlinear optical elements, such as optical parametric amplifiers (OPAs), which can generate squeezed states to boost quantum correlations, to further enhance entanglement in molecular COM systems remains a largely unexplored territory, despite their established benefits in conventional setups~\cite{Zhang2018,Pan2022,Kibret2023}. This is a critical oversight, as molecular COM systems exhibit unique collective molecular modes and nonlinear optomechanical interactions that could fundamentally alter the impact of the OPA.
	
	To address this knowledge gap, we present a rigorous theoretical analysis of a double-cavity molecular optomechanical system incorporating an intracavity OPA, focusing on its nuanced influence on bipartite entanglement. Importantly, and to the best of our knowledge for molecular COM systems, we systematically analyze the OPA's differential impact on entanglement between distinct degrees of freedom: cavity photons and molecular vibrations (optical-vibration entanglement), essential for quantum state transfer and readout, and between vibrational modes themselves (vibration-vibration entanglement), a critical resource for quantum memories and transducers. Our key findings reveal a counterintuitive trade-off: while the OPA significantly enhances the vibration-vibration entanglement (by up to 40\% under optimal conditions), it simultaneously attenuates the optical-vibration entanglement. Furthermore, we demonstrate that maximal vibration-vibration entanglement is achieved under conditions of symmetric population of the molecular collective modes, a finding consistent with recent theoretical predictions for hybrid quantum systems~\cite{Huang2024} and directly relevant for experimental optimization.
	
	By elucidating the intricate role of OPAs in shaping entanglement within molecular COM systems, our work provides a critical step towards realizing the quantum technological potential of these platforms. Our theoretical framework offers a pathway to optimized entanglement generation and manipulation, exploiting the advantageous room-temperature operation and scalability of molecular COM systems. The demonstrated enhancement of vibration-vibration entanglement, coupled with the inherent thermal robustness we will showcase, positions this approach as highly promising for applications in quantum computing, quantum communication, and advanced quantum sensing.
	
	The remainder of this paper is structured as follows. In \Cref{sec:Model} we provide a detailed description of our model, i.e., a molecular cavity optomechanical system that incorporates $N$ molecules, accompanied by the formulation of the Hamiltonian and the derivation of the equations of motion. Furthermore, the output field of the system is calculated. \Cref{sec:Resul} delves into the achievement of bipartite entanglement through numerical analysis. Finally, \Cref{sec:Concl} summarizes our key findings, discusses their implications, and outlines promising avenues for future research.
	
	\section{Model and dynamics}\label{sec:Model}
	
	\subsection{Hamiltonian of the system}\label{sec:Hamil}
	
\begin{figure}[htp!]
		\setlength{\lineskip}{0pt}
		\centering
		{\includegraphics[width=9.1cm]{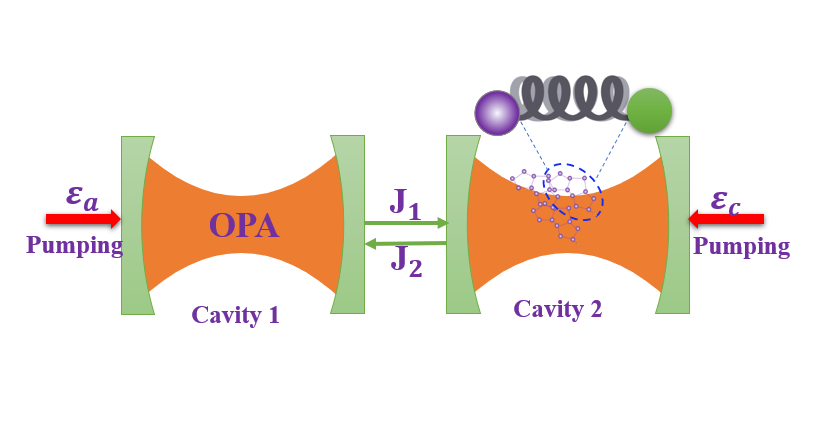}}
		\caption{ Schematic diagram of the molecular optomechanical system, consisting of Cavity 1 (Fabry-Pérot) hosting an optical parametric amplifier (OPA) and Cavity 2 (plasmonic) containing N identical molecules. The cavities are coupled in a nonreciprocal manner, establishing a one-way interaction, indicated by directional coupling strengths $J_1$ and $J_2$. Pumping fields $(\mathcal{E}_a,\mathcal{E}_c)$ drive both cavities.}
		\label{fig:fig1}
	\end{figure}

	Our benchmark system, depicted schematically in \Cref{fig:fig1}, is a molecular cavity optomechanical (McOM) system. It consists of a Fabry-Pérot cavity, which we designate as Cavity 1, incorporating an optical parametric amplifier (OPA), and a separate plasmonic cavity, Cavity 2, containing $N$ identical molecules. As detailed in \textcite{Roelli2024}, plasmonic cavities, such as Cavity 2, are essential to achieve strong coupling in McOM systems due to their ability to confine light to nanoscale volumes, enhancing light-matter interactions. In our system, the second cavity mode corresponds to the plasmonic mode of a metallic nanoparticle~\cite{esteban2022molecular}.
	
	We consider strong pumping fields with amplitudes $\mathcal{E}_j$ and frequencies $\omega_{0j}$ to directly drive both cavities ($j=1,2$), with molecular vibration modeled as a quantum mechanical oscillator of frequency $\omega_m$. We assume the driving fields are phase-locked. The Hamiltonian is derived in a frame rotating at the laser frequency for cavity modes ($\omega_{01}$ and $\omega_{02}$), and in a frame rotating at the molecular vibration frequency $\omega_m$ for the vibrational modes. For simplicity in the subsequent linearization, we assume $\omega_{01} = \omega_{02} = \omega_{\rm drive}$. The system Hamiltonian, within the rotating-wave approximation and setting $\hbar=1$, is (see \Cref{sec:Ham} for derivation details):
	\begin{equation}\label{eq: Hamil1}
	\begin{aligned}
	\tilde{H}_{\mathrm{new}}=&\Delta_a a^\dagger a+\Delta_{c} c^\dagger c  +\sum_{j=1}^N \omega_m b_j^\dagger b_j\\
	&-\sum_{j=1}^N g_m c^\dagger c(b_j^\dagger + b_j)+J_1a^\dagger c+J_2c^\dagger a \\
	&+i\Lambda(e^{i\theta}a^{\dagger 2}\tcb{}-e^{-i\theta}a^2\tcb{})\\ 
	&+ i\mathcal{E}_a(a^\dagger - a)+i\mathcal{E}_c(c^\dagger - c).
	\end{aligned}
	\end{equation}
	Here, $a(a^\dagger)$ and $c(c^\dagger)$ are the annihilation (creation) operators for cavity fields 1 and 2 respectively, with resonance frequencies $\omega_a$ and $\omega_c$. $b_j(b^\dagger_j)$ are the annihilation (creation) operators for the $j^{th}$ molecular vibration mode, satisfying bosonic commutation relations $[\mathcal{O},O^\dagger]=1$. The Hamiltonian terms describe: cavity mode energies ($\Delta_a a^\dagger a, \Delta_{c} c^\dagger c$), molecular vibrational energy ($\sum_{j=1}^N \omega_m b_j^\dagger b_j$), Cavity 2 - molecular vibration interaction ($\sum_{j=1}^N g_m c^\dagger c(b_j^\dagger + b_j)$), nonreciprocal inter-cavity coupling ($J_1a^\dagger c+J_2c^\dagger a$). We introduce non-reciprocal coupling ($J_1 \ne J_2$) to explore its potential for enhancing entanglement by preventing destructive interference effects that can occur in reciprocally coupled systems, as demonstrated in \Cref{sec:results_nonreciprocal}. OPA interaction within Cavity 1 ($i\Lambda(e^{i\theta}a^{\dagger 2}\tcb{}-e^{-i\theta}a^2\tcb{})$), and driving fields ($i\mathcal{E}_a(a^\dagger - a)+i\mathcal{E}_c(c^\dagger - c)$). $\Delta_a=\omega_a-\omega_{01}$ and $\Delta_c=\omega_c-\omega_{02}$ represent the detunings of the pumping fields from their respective cavity modes.
	
	To investigate collective effects and facilitate the analysis of entanglement, we introduce collective vibrational modes:
	\begin{align}\label{eq: CollectiveModes}
	&B_1=\sum_{j=1}^{M}\frac{b_j}{\sqrt{M}}, &B_2=\sum_{j=M+1}^{N}\frac{b_j}{\sqrt{N-M}}.
	\end{align}
	Here, $B_1$ and $B_2$ represent collective vibration modes composed of $M$ and $N-M$ molecules, respectively, where $N$ is the total number of molecules in cavity 2 and $M$ is the number of molecules included in the $B_1$ collective mode. This theoretical partitioning allows us to investigate how dividing the molecular ensemble affects the entanglement between different collective modes. $M$ is a theoretical parameter ranging from $0$ to $N$, used to explore optimal entanglement configurations as shown in \Cref{fig:fig5}. Experimental implementations might involve spatially separating molecular subgroups or addressing distinct vibrational modes to realize similar partitioning. In terms of these collective modes, the Hamiltonian becomes:
	\begin{equation}\label{eq:Hamil2}
	\begin{aligned}
	\tilde{H}_{\mathrm{new}}=&\Delta_a a^\dagger a+\Delta_{c} c^\dagger c  +\sum_{k=1,2}\omega_m B_k^\dagger B_k\\
	&-\sum_{k=1,2}\left(g_k c^\dagger c(B_k^\dagger + B_k)\right)+J_1a^\dagger c+J_2c^\dagger a\\ &+i\Lambda(e^{i\theta}a^{\dagger 2}\tcb{}-e^{-i\theta}a^2\tcb{})\\ 
	&+ i\mathcal{E}_a(a^\dagger - a)+i\mathcal{E}_c(c^\dagger - c),
	\end{aligned}
	\end{equation}
	where $g_1=g_m\sqrt{M}$ and $g_2=g_m\sqrt{N-M}$ are the collective optomechanical coupling strengths between cavity mode $c$ and collective vibrational modes $B_1$ and $B_2$. As emphasized by \textcite{Huang2024}, the use of collective operators is pivotal for capturing the enhanced coupling and entanglement in McOM systems with large numbers of molecules.
	
	While our theoretical framework demonstrates the potential for entanglement enhancement, practical implementation requires addressing several challenges: (i) Maintaining balanced gain/loss ratios in molecular systems while preserving strong optomechanical coupling; (ii) Precision control of coupling modulation frequencies ($\omega_m$) to maintain exceptional point configurations; (iii) Thermal stability of plasmonic cavity parameters under strong pumping conditions. Recent advances in nanofabrication techniques \cite{Roelli2024} and active control of exceptional points in optomechanical arrays \cite{Huang2024} suggest promising paths for experimental realization. The implementation of the LC circuit demonstrated in \cite{Chitsazi2017} provides an alternative platform for testing these concepts prior to molecular implementation.
	
	\subsection{Quantum Langevin Equations}\label{sec:QLE}
	
	To analyze the dynamics of this open quantum system, we employ the standard tool of quantum Langevin equations (QLEs), which incorporate quantum noise and dissipation from the environment. These equations are derived from the time-independent Hamiltonian in \Cref{sec:Hamil} by coupling the system to thermal reservoirs and applying the input-output formalism \cite{Gardiner2004}. The equations are written in a frame rotating at the driving frequency for the cavity modes and in a frame rotating at the molecular vibration frequency $\omega_m$ for the vibrational modes.
	\begin{equation}\label{eq:QLE1}
	\begin{aligned}
	\dot{a}=&-(i\Delta_{a}+\kappa_a)a-iJ_1c+2\Lambda e^{i\theta}a^{\dagger}+\mathcal{E}_a+\sqrt{2\kappa_a}a^{in},\\
	\dot{c}=&-(i\Delta_c+\kappa_c)c+ic\sum_{k=1}^{2}g_k(B_k+B_k^\dagger)-iJ_2a\\
	&+\mathcal{E}_c+\sqrt{2\kappa_c}c^{in},\\
	\dot{B}_1=&-(i\omega_m+\gamma_1)B_1+ig_1c^\dagger c+\sqrt{2\gamma_1}B_1^{in},\\
	\dot{B}_2=&-(i\omega_m+\gamma_2)B_2+ig_2c^\dagger c+\sqrt{2\gamma_2}B_2^{in}.
	\end{aligned}
	\end{equation}
	Here, $\kappa_a$ and $\kappa_c$ are the energy decay rates for cavities 1 and 2, respectively, describing the loss of photons in the cavity to the environment. $\gamma_1$ and $\gamma_2$ are the decay rates for collective vibrational modes $B_1$ and $B_2$, representing the dissipation of vibrational energy into a thermal bath. $\mathcal{O}^{\text{in}} (\mathcal{O}=a,c,B_k)$ represent input noise operators, representing the influx of quantum noise from the environment into the system modes, which, for collective vibration modes, are defined as:
	\begin{align}\label{eq:NoiseModes}
	&B^{\text{in}}_1=\frac{1}{\sqrt{M}}\sum_{j=1}^{M}b^{\text{in}}_j, & B^{\text{in}}_2=\frac{1}{\sqrt{N-M}}\sum_{j=M+1}^{N}b^{\text{in}}_j.
	\end{align}
	These noise operators are assumed to have zero mean values and are characterized by the following correlation functions, consistent with the standard quantum noise formalism for open quantum systems~\cite{Gardiner2004}:
	\begin{equation}\label{eq:NoiseCorrelations}
	\begin{aligned}
	&\langle a^{\text{in}}(t)a^{\text{in}\dagger}(t^\prime)\rangle=\delta(t-t^\prime), \\
	& \langle c^{\text{in}}(t)c^{\text{in}\dagger}(t^\prime)\rangle=\delta(t-t^\prime),\\
	&\langle B_k^{\text{in}}(t)B_k^{\text{in}\dagger}(t^\prime)\rangle=(n_{\text{th},k}+1)\delta(t-t^\prime), \\
	& \langle B_k^{\text{in}\dagger}(t)B_k^{\text{in}}(t^\prime)\rangle=n_{\text{th},k}\delta(t-t^\prime),
	\end{aligned}
	\end{equation}
	where $n_{\text{th},k}=\left\{\exp\left(\frac{\hbar\omega_m}{k_BT}\right)-1\right\}^{-1}$ is the thermal phonon occupation number from Bose-Einstein statistics at temperature $T$. Note that since both collective modes $B_1$ and $B_2$ are at frequency $\omega_m$, $n_{\text{th},1}=n_{\text{th},2}=n_{\text{th}}$. and $k_B$ is the Boltzmann constant. These correlation functions reflect the fundamental quantum nature of the noise, being delta-correlated in time and having variances determined by the quantum statistics of the reservoirs.
	
	To analyze the steady-state behavior and quantum entanglement of the system, we linearize the non-linear QLEs in \Cref{eq:QLE1} by expressing each operator as a sum of its steady-state mean value and a small fluctuation around it: $a=\alpha_a+\delta a$, $c=\alpha_c+\delta c$, $B_k=\beta_k+\delta B_k$. This linearization, valid under strong driving conditions, allows us to analyze quantum fluctuations around the steady state. The steady-state solutions, representing the mean-field amplitudes of the cavity and vibrational modes under continuous driving, are found to be the following:
	\begin{equation}\label{eq:SteadyState}
	\begin{aligned}
	\alpha_a&=\frac{\mathcal{E}_a-iJ_1\alpha_c}{(i\Delta_a+\kappa_a)-2\Lambda e^{i\theta}}, \\
	\alpha_c&=\frac{\mathcal{E}_c-iJ_2\alpha_a}{(i\tilde{\Delta}_c+\kappa_c)}, \\
	\beta_k&=\frac{-ig_k|\alpha_c|^2}{(i\omega_m+\gamma_k)},
	\end{aligned}
	\end{equation}
	where $\tilde{\Delta}_c=\Delta_c-\sum_{k=1}^{2}2g_k\text{Re}[\beta_k]$ is the normalized driving detuning, and we define the linearized coupling strengths $G_1=g_1|\alpha_c|$ and $G_2=g_2|\alpha_c|$. The steady-state amplitudes $\alpha_a$ and $\alpha_c$ are generally complex, as shown by their definitions in \Cref{eq:SteadyState}. For simplicity in the subsequent analysis, we assume real steady-state amplitudes $\alpha_a$ and $\alpha_c$ by appropriately choosing the phase reference of the cavity fields.
	We note that this simplification cannot be achieved by adjusting the phase of the driving fields alone, due to the complex structure of the self-consistent equations (see \Cref{eq:SteadyState}). The choice of a real $\alpha_c$  should therefore be viewed as a specific phase convention that simplifies the linearized analysis without loss of generality for the physical predictions considered in this work.
	The analysis of quantum fluctuations around the steady state is valid regardless of whether the steady-state amplitudes are real or complex, as long as \Cref{eq:SteadyState} represents the correct steady-state solution and the linearization condition is met. The assumption of real amplitudes simplifies the form of the drift matrix $\mathcal{A}$. Assuming identical decay rates for both collective vibrational modes ($\gamma_1=\gamma_2=\gamma$) and cavity modes ($\kappa_a=\kappa_c=\kappa$), and neglecting thermal excitation in the optical frequency band, the linearized quantum Langevin equations for the fluctuation operators become the following:
	\begin{equation}\label{eq:Fluc}
	\begin{aligned}
	\delta\dot{a}=&-(i\Delta_a+\kappa_a)\delta a-iJ_1\delta c+2\Lambda e^{i\theta}\delta a^\dagger+\sqrt{2\kappa_a}a^{in},\\
	\delta\dot{c}=&-(i\tilde{\Delta}_c+\kappa_c)\delta c+iG_1(\delta B_1+\delta B_1^\dagger)\\&+iG_2 (\delta B_2+\delta B_2^\dagger)-iJ_2\delta a+\sqrt{2\kappa_c}c^{in},\\
	\delta\dot{B}_1=&-(i\omega_m+\gamma_1)\delta B_1+iG_1(\delta c+\delta c^\dagger)+\sqrt{2\gamma_1}B_1^{in},\\
	\delta\dot{B}_2=&-(i\omega_m+\gamma_2)\delta B_2+iG_2(\delta c+\delta c^\dagger)+\sqrt{2\gamma_2}B_2^{in}.
	\end{aligned}
	\end{equation}
	
	To analyze entanglement, we transform them into quadrature operators for the optical and vibrational modes:
	\begin{equation}\label{eq:Quadratures}
	\begin{aligned}
	\delta x_{l=a,c}&=\frac{\delta \mathcal{O}+\delta \mathcal{O}^{\dagger}}{\sqrt2},
	& \delta y_{l=a,c}=\frac{\delta \mathcal{O}- \delta \mathcal{O}^{\dagger}}{i\sqrt2},\\
	\delta q_k&=\frac{\delta B_k+\delta B_k^{\dagger}}{\sqrt2},
	& \delta p_k=\frac{\delta B_k-\delta B_k^{\dagger}}{i\sqrt2},
	\end{aligned}
	\end{equation}
	and their corresponding input noise quadratures:
	\begin{equation}\label{eq:InputQuadratures}
	\begin{aligned}
	x_{l=a,c}^{\text{in}}&=\frac{\mathcal{O}^{\text{in}}+ \mathcal{O}^{\text{in}\dagger}}{\sqrt2},
	& y_{l=a,c}^{\text{in}}=\frac{ \mathcal{O}^{\text{in}}- \mathcal{O}^{in\dagger}}{i\sqrt2},\\
	q^{\text{in}}_k&=\frac{ B_k^{\text{in}}+ B_k^{\text{in}\dagger}}{\sqrt2},
	& p_k^{\text{in}}=\frac{ B_k^{\text{in}}- B_k^{\text{in}\dagger}}{\sqrt2},
	\end{aligned}
	\end{equation}
	where $\mathcal{O}^{\text{in}} (\mathcal{O}=a,c,B_k)$ and $k=1,2$. Using these quadratures, the linearized QLEs (\Cref{eq:Fluc}) can be written in a compact matrix form:
	\begin{equation}\label{eq:MatrixForm}
	\dot{\Gamma}(t)=\mathcal{A}\Gamma(t)+n(t),
	\end{equation}
	where the drift matrix yields,
	\begin{widetext}
		\begin{equation}\label{eq:DriftMatrix}
		\mathcal{A}=
		\begin{pmatrix}
		2\Lambda\cos\theta-\kappa_a&2\Lambda\sin\theta+\Delta_a&0&J_1&0&0&0&0\\
		2\Lambda\sin\theta-\Delta_a&-(2\Lambda\cos\theta+\kappa_a)&-J_1&0&0&0&0&0\\
		0&J_2&-\kappa_{c}&\tilde{\Delta}_c&0&0&0&0\\
		-J_2&0&-\tilde{\Delta}_c&-\kappa_{c}&2G_1&0&2{G}_2&0\\
		0&0&0&0&-\gamma_1&\omega_{m}&0&0\\
		0&0&2G_1&0&-\omega_{m}&-\gamma_1&0&0\\
		0&0&0&0&0&0&-\gamma_2&\omega_{m}\\
		0&0&2G_2&0&0&0&-\omega_{m}&-\gamma_2\\
		\end{pmatrix},
		\end{equation}
		with the vector of the quadrature fluctuation operators,
		\begin{equation}\label{eq:FluctuationVector}
		\Gamma(t)=\left(\delta x_{a}(t),\delta y_{a}(t),\delta x_{c}(t),\delta y_{c}(t),\delta q_1(t),\delta p_1(t),\delta q_2(t),\delta p_2(t)\right)^{T},
		\end{equation}
		and the vector of the noise operators reads,
		\begin{equation}\label{eq:NoiseVector}
		n(t)=\left(\sqrt{2\kappa_a}x_{a}^{\text{in}},\sqrt{2\kappa_a}y_{a}^{\text{in}},\sqrt{2\kappa_c}x_{c}^{\text{in}},\sqrt{2\kappa_c}y_{c}^{\text{in}},\sqrt{2\gamma_1}q_1^{\text{in}},\sqrt{2\gamma_1}p_1^{\text{in}},\sqrt{2\gamma_2}q_2^{\text{in}}, \sqrt{2\gamma_2}p_2^{\text{in}}\right)^{T}.
		\end{equation}
	\end{widetext}
	
	The stability of the system is determined by the eigenvalues of the drift matrix $\mathcal{A}$. Stability requires that all real parts of the eigenvalues of $\mathcal{A}$ are negative, a condition that we analyze numerically using the Routh-Hurwitz criterion~\cite{DeJesus1987}. Due to the Gaussian nature of quantum noise and the linearity of QLEs, the quantum state of the system is fully characterized by the $8 \times 8$ covariance matrix (CM) $V$, which is obtained by solving the Lyapunov equation:
	\begin{equation}\label{eq:Lyapunov}
	\mathcal{A}V+V\mathcal{A}^{T}=-D,
	\end{equation}
	where $V_{ij}=\frac{1}{2}\left\{\Gamma_{i}(t)\Gamma_{j}(t^{\prime})+\Gamma_{j}(t^{\prime})\Gamma_{i}(t)\right\}$, and the diffusion matrix $D$ is:
	\begin{widetext}
		\begin{equation}\label{eq:DiffusionMatrix}
		D=\text{diag}\left[\kappa_{a},\kappa_{a},\kappa_{c},\kappa_{c},\gamma_1(2n_{\text{th}}+1),\gamma_1(2n_{\text{th}}+1),\gamma_2(2n_{\text{th}}+1),\gamma_2(2n_{\text{th}}+1)\right],
		\end{equation}
		and is defined through $\langle v_{k}(t)v_{l}(t^\prime)+v_{l}(t^\prime)v_{k}(t)\rangle/2=D_{kl}\delta(t-t^\prime)$. The covariance matrix $V$ is then used to quantify the bipartite entanglement.
	\end{widetext}
	
	\subsection{Quantification of Bipartite Entanglement}\label{sec:Quantif}
	
	To quantify bipartite entanglement, we employ logarithmic negativity $E^N$, a well-established measure for continuous-variable Gaussian states~\cite{Plenio2005}. The logarithmic negativity is defined as:
	\begin{equation}
	E^N=\max[0,-\ln(2\zeta)],
	\end{equation}
	where $\zeta = 2^{-1/2} \{\Sigma(V) - [\Sigma(V)^2 - 4\text{det}(V)]^{1/2}\}^{1/2}$ and $\Sigma(V) = \text{det}(\varPhi_1)+\text{det}(\varPhi_2)-2\text{det}(\varPhi_3)$. Here, $\varPhi_{i}$ are sub-matrices derived from the covariance matrix $V$ for a given bipartite subsystem. For a subsystem composed of modes 1 and 2, the covariance matrix can be partitioned as:
	\begin{equation}\label{eq:CovariancePartition}
	V_{\text sub}=
	\begin{pmatrix}
	\varPhi_1 & \varPhi_3 \\
	\varPhi_3^{T} & \varPhi_2 \\
	\end{pmatrix}.
	\end{equation}
	Throughout this work, we focus on quantifying the bipartite entanglement between cavity mode $c$ and vibrational modes $B_{1,2}$ (denoted $E_{cB_j}^N$) and between the two vibrational modes $B_1$ and $B_2$ (denoted $E_{B_1B_2}^N$). We present results for $E_{cB_2}^N$ and $E_{B_1B_2}^N$ as representative cases, noting that the optical-vibration entanglement $E_{cB_1}^N$ exhibits qualitatively similar behavior.
	
	\section{Results and discussion}\label{sec:Resul} 
	
	In this section, we present and discuss the results of our numerical analysis, exploring the stability and bipartite entanglement properties of the proposed McOM system. Our analysis is based on numerical simulations that employ experimentally feasible parameters, selected based on previous experimental and theoretical studies of McOM and related systems~\cite{Xiao2012,Liu2017,Zou2024,Schmidt2024,Huang2024}: $\omega_m/2\pi = \SI{30}{\tera\hertz}, \quad g_m/2\pi = \SI{30}{\giga\hertz}, \quad \kappa_a/\omega_m = 0.3, \quad \kappa_a=\kappa_c, \quad \Delta_a=\omega_m, \quad \Delta_c=\omega_m, \quad \gamma_1/\omega_m = \num{1e-4}, \quad \gamma_2=\gamma_1, \quad \frac{\mathcal{E}}{\omega_m} = 16, \quad J_1/\omega_m = 0.9, \quad J_2/\omega_m = 0.3, \quad M=50, \quad N=\num{100}, \quad \theta=\frac\pi2 \quad \text{and} \quad T=\SI{312}{\kelvin}$. These parameters, particularly the high vibrational frequency $\omega_m$ and the large optomechanical coupling $g_m$, are characteristic of molecular COM systems and are chosen to ensure experimental relevance, reflecting current technological capabilities in the field~\cite{Roelli2024,schmidt2017}.
	
	The parameters used in our simulations are chosen to be in line with the values reported in the recent literature on molecular cavity optomechanics and related hybrid systems \cite{Roelli2024,Zou2024,Schmidt2024,Huang2024}. The high vibrational frequency $\omega_m$ and the large optomechanical coupling $g_m$ highlight the key advantages of McOM systems. We note that achieving some specific parameters simultaneously in a single experimental setup presents significant technological challenges. For instance, while engineered non-reciprocal coupling has been explored in other photonic platforms, its realization in nanoscale plasmonic cavities is an ongoing challenge. Similarly, while the mechanical dissipation rate $\gamma/\omega_m = \num{1e-4}$ ($\gamma/2\pi \approx \SI{3}{\giga\hertz}$) represents a relatively low-dissipation scenario for optical phonons, such values are theoretically plausible in certain engineered environments or at lower temperatures. The driving amplitude $\mathcal{E}/\omega_m=16$ corresponds to a strong driving field; the resulting steady-state photon numbers $|\alpha_a|^2$ and $|\alpha_c|^2$ would depend on the specific decay rates and detunings of the cavity, and achieving very high photon numbers can be experimentally demanding. The range of OPA gain $\Lambda/\omega_m$ up to $0.4$ is motivated by theoretical and experimental studies of OPAs in optical cavities relevant for squeezing and parametric effects \cite{Zhang2018,Pan2022,Kibret2023}. Although experimental realization of our proposed scheme would require overcoming these challenges, our theoretical investigation aims to explore the potential of McOM systems under such conditions, providing a roadmap for future experimental efforts.
	
	\subsection{Stability of the molecular optomechanical system}\label{sec:stability} 
	
\begin{figure}[htp!]
		\centering
		\includegraphics[width=1.05\linewidth]{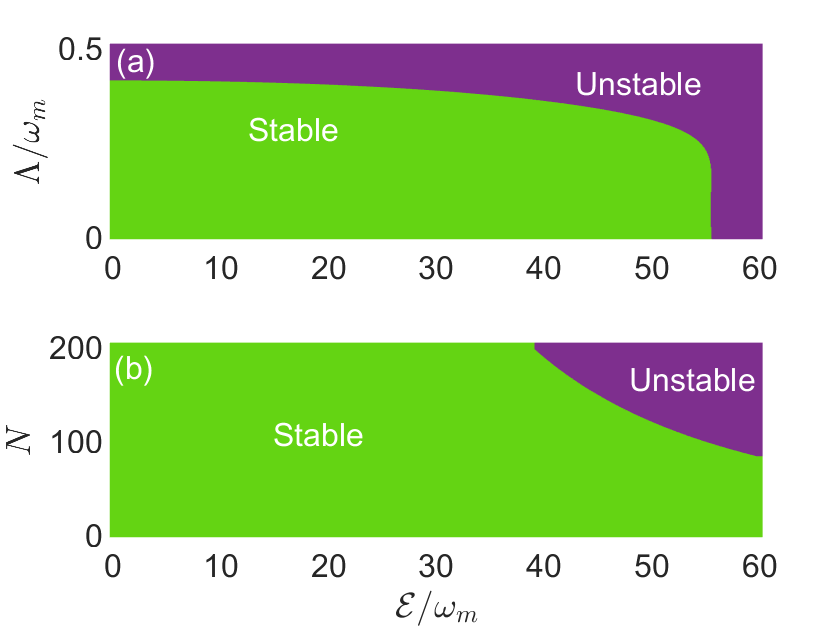}
		\caption{System stability analysis as a function of driving amplitude $\mathcal{E}$ and (a) nonlinear gain of OPA $\Lambda$ (b) number of molecules $N$. Parameters are: $\omega_m/2\pi=\SI{30}{\tera\hertz}$, $g_m/2\pi=\SI{30}{\giga\hertz}$, $\kappa/\omega_m=0.3$, $\Delta_a=\omega_m$, $\Delta_c=\omega_m$, $\gamma_1/\omega_m=\num{1e-4}$, $\gamma_2=\gamma_1$, $J_1/\omega_m=0.9$, $J_2/\omega_m=0.3$, $M=50$, $\theta=\frac\pi2$, and $T=\SI{312}{\kelvin}$. In (a), $N=\num{100}$ and in (b), $\frac{\Lambda}{\omega_m}=0.2$. Stable (green) and unstable (purple) regions are clearly delineated, indicating parameter regimes for achieving steady-state entanglement. Note that these stability boundaries are specific to the fixed values of the other parameters listed.}
		\label{fig:fig2}
	\end{figure}

	Before investigating entanglement, it is essential to establish the parameter regimes for stable system operation. This subsection presents a stability analysis of our McOM system as a function of key control parameters. \Cref{fig:fig2} displays the stability regions as a function of the driving amplitude $\mathcal{E}$ and the non-linear gain of OPA $\Lambda$ (\Cref{fig:fig2}a) and the number of molecules $N$ (\Cref{fig:fig2}b). Stability is assessed by ensuring that all eigenvalues of the drift matrix $\mathcal{A}$ have negative real parts, using the Routh-Hurwitz criterion~\cite{DeJesus1987}.
	
	As shown in \Cref{fig:fig2}(a), the system exhibits stability within a defined parameter space, bounded by $0\leq\frac{\mathcal{E}}{\omega_m}\leq55.4$ and $0\leq\frac{\Lambda}{\omega_m}\leq0.4$. In particular, increasing the OPA gain $\Lambda$ above a threshold ($\frac{\Lambda}{\omega_m} > 0.4$) drives the system into an unstable regime. This instability arises from the parametric amplification process inherent to OPAs: high nonlinear gain amplifies quantum fluctuations, leading to an exponential growth of system oscillations and a transition to instability. Similarly, \Cref{fig:fig2}(b) reveals that increasing the number of molecules to $N>\num{100}$ also destabilizes the system at higher driving amplitudes $\mathcal{E}$. This destabilization with increasing $N$ can be attributed to the enhanced collective coupling strength, which intensifies nonlinear interactions and further amplifies fluctuations. Experimentally, maintaining stability beyond these parameter regimes would necessitate reducing the driving amplitude to counteract the enhanced nonlinear response, particularly for larger molecular ensembles.
	
	The significance of this stability analysis lies in its role as a practical guide for experimentalists aiming to harness quantum entanglement in molecular optomechanics. By delineating stable parameter regimes, it enables researchers to optimize entanglement generation while avoiding instability. In particular, these stability boundaries are specific to the chosen parameters, such as low mechanical dissipation and nonreciprocal coupling; in setups with higher dissipation or reciprocal coupling, the boundaries can shift, necessitating further studies with realistic parameters to ensure practical applicability.
	
	\subsection{Entanglement enhancement via OPA gain and driving amplitude}\label{sec:OPA_ent} 
	
	\begin{figure}[htp!]
		\centering
		\includegraphics[width=9.1cm]{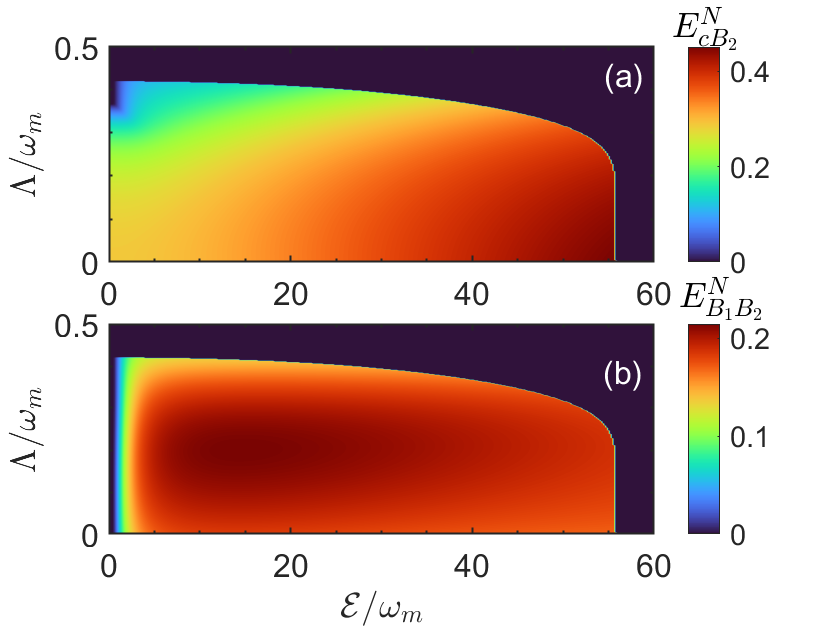}	
		\caption{Contour plots of bipartite entanglement as a function of driving amplitude $\frac{\mathcal{E}}{\omega_m}$ and OPA nonlinear gain $\Lambda/\omega_{m}$ (for $N=\num{100}$ molecules). (a) optical-vibration entanglement ($E_{cB_2}^N$) and (b) Vibration-vibration entanglement ($E_{B_1B_2}^N$). Entanglement is quantified by Logarithmic Negativity. Parameters are identical to \Cref{fig:fig2}. Entanglement is observed to be achievable only within the stable parameter region identified in \Cref{fig:fig2}.}
		\label{fig:fig3}
	\end{figure}
	
	Having established the stability boundaries, we now investigate the bipartite entanglement within the stable regime. \Cref{fig:fig3} presents contour plots of (a) optical-vibration entanglement ($E_{cB_2}^N$) and (b) vibration-vibration entanglement ($E_{B_1B_2}^N$) as functions of driving amplitude $\frac{\mathcal{E}}{\omega_m}$ and nonlinear OPA gain $\frac{\Lambda}{\omega_m}$. Consistent with the stability analysis, steady-state entanglement is achievable only within the stable parameter region (i.e., $\frac{\Lambda}{\omega_m} < 0.4$ and $\frac{\mathcal{E}}{\omega_m}\lesssim 55.4$). In particular, \Cref{fig:fig3}(b) shows that vibration-vibration entanglement ($E_{B_1B_2}^N$) is negligible at very low driving amplitudes, indicating a requirement for sufficient driving power to induce entanglement between the vibration modes. Similarly, the optical-vibration entanglement ($E_{cB_2}^N$) is also weak at low driving amplitudes (\Cref{fig:fig3}(a)). This observation is consistent with the underlying physics: generating significant entanglement requires a sufficiently strong optomechanical coupling, which, 
	in our system, is enhanced by higher driving fields, leading to a greater effective coupling strength $G_2$.
	
	\subsection{Entanglement scaling with molecule number and saturation effects}\label{sec:scaling}
	
	\begin{figure}[htp!]
		\centering
		\includegraphics[width=9.1cm]{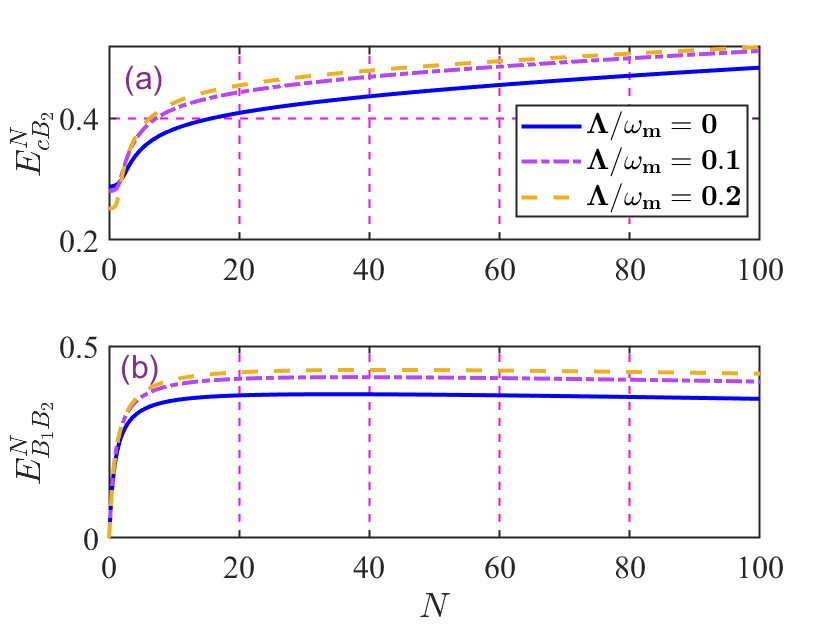}	
		\caption{Entanglement scaling with molecule number $N$. (a) optical-vibration entanglement $E^N_{cB_2}$ and (b) vibration-vibration entanglement $E_{B_1B_2}^N$ as a function of molecule number $N$ for different OPA nonlinear gains $\frac{\Lambda}{\omega_m}$. Entanglement is quantified by Logarithmic Negativity. Parameters are the same as \Cref{fig:fig2}, except $M=0$ and $\frac{\mathcal{E}}{\omega_m}=16$. The optical-vibration entanglement shows monotonic increase with $N$, while vibration-vibration entanglement plateaus rapidly.}
		\label{fig:fig4}
	\end{figure} 
	
	To understand how the collective nature of the molecular ensemble affects the entanglement, we next examine the scaling of the entanglement with the number of molecules $N$. \Cref{fig:fig4} explores this scaling, first focusing on the optical-vibration entanglement. \Cref{fig:fig4}(a) shows the optical-vibration entanglement $E^N_{cB_2}$ as a function of $N$ for various OPA gains $\frac{\Lambda}{\omega_m}$, specifically when the collective mode distribution number $M=0$. 
	
	In this configuration ($M=0$), the collective coupling strength $G_2 = g_m\sqrt{N}|\alpha_c|$ scales with $\sqrt{N}$, indicating that increasing the number of molecules enhances the optical-vibration coupling. In fact, we observe that $E^N_{cB_2}$ increases with $N$, reaching a maximum value of $E^N_{cB_2}\approx0.51$ when the OPA is active ($\frac{\Lambda}{\omega_m} > 0$). The physical origin of this enhancement can be attributed to the increased collective coupling strength and the OPA-driven amplification of quantum correlations. Interestingly, beyond a certain number of molecules, the entanglement saturates, suggesting a limit to the benefit of simply increasing $N$ in this configuration~\cite{Djorwe2022}. This saturation may be related to increased losses or dephasing at higher molecular densities, effects not explicitly included in our simplified model but relevant in realistic McOM systems.
	
	Complementing the analysis of optical-vibration entanglement, we now examine the scaling of vibration-vibration entanglement with molecule number and explore the role of molecular distribution between collective modes. \Cref{fig:fig4}(b) presents the vibration-vibration entanglement $E_{B_1B_2}^N$ as a function of $N$. 
	
	\Cref{fig:fig4}(b) plots the vibration-vibration entanglement $E_{B_1B_2}^N$ as a function of $N$. In contrast to optical-vibration entanglement, $E_{B_1B_2}^N$ exhibits a non-monotonic dependence on $N$. Although OPA continues to improve $E_{B_1B_2}^N$, the entanglement level plateaus rapidly with increasing $N$. This behavior suggests that vibration-vibration entanglement is less directly enhanced by increasing the total number of molecules beyond a certain point and that other factors, such as the distribution of molecules between collective modes, may play a more dominant role.
	
	\begin{figure}[htp!]
		\centering
		\includegraphics[width=9.1cm]{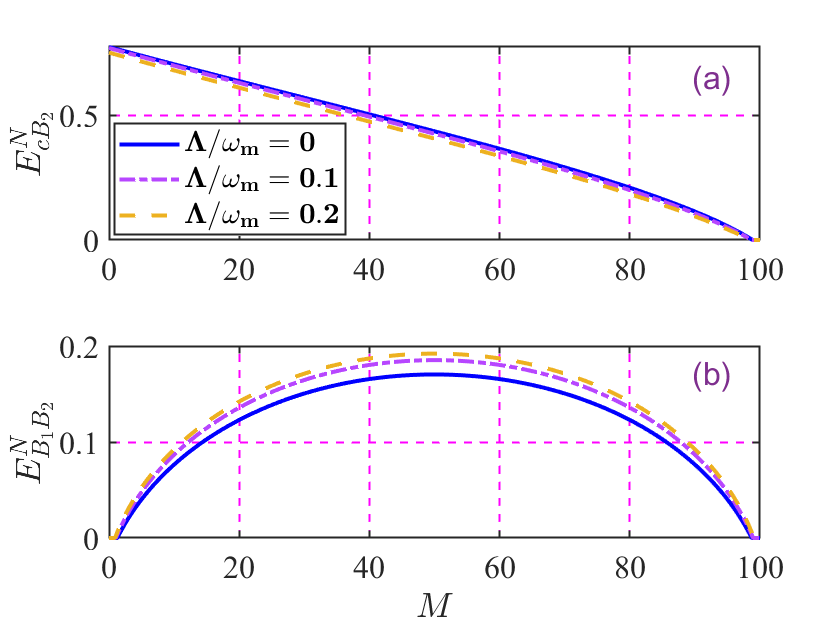}	
		\caption{Bipartite entanglement as a function of molecular collective mode distribution $M$ (for $N=\num{100}$ molecules and $\frac{\mathcal{E}}{\omega_m}=50$). (a) optical-vibration entanglement $E^N_{cB_2}$ and  (b) vibration-vibration entanglement $E_{B_1B_2}^N$. Entanglement is quantified by Logarithmic Negativity. Parameters are identical to \Cref{fig:fig2}. Vibration-vibration entanglement peaks when the molecular population is symmetrically distributed ($M=\frac N2$), while optical-vibration entanglement is maximized for $M=0$. Note that vibration-vibration entanglement is zero when $M=0$ or $M=N$ by definition of the collective modes (Eq. \eqref{eq:CollectiveModes}).}
		\label{fig:fig5}
	\end{figure}
	
	To further explore the role of collective modes, \Cref{fig:fig5} examines the bipartite entanglement as a function of the distribution number $M$ of the molecular collective mode $B_1$, while keeping the total number of molecules constant ($N=\num{100}$). \Cref{fig:fig5}(a) reveals that optical-vibration entanglement $E^N_{cB_2}$ is maximized when $M=0$ and decreases monotonically as $M$ increases. In contrast, the vibration-vibration entanglement $E_{B_1B_2}^N$ (\Cref{fig:fig5}(b)) exhibits a strikingly different behavior: it is zero when $M=0$ or $M=N$, and reaches its maximum value when $M=\frac N2=50$. This indicates that vibration-vibration entanglement is maximized when the molecules are equally distributed between the two collective modes.
	
	This dependence on $M$ can be understood by considering the effective optomechanical coupling strengths $G_1 = g_m\sqrt{M}|\alpha_c|$ and $G_2 = g_m\sqrt{N-M}|\alpha_c|$. When $M=0$, $G_1=0$, effectively decoupling the collective mode $B_1$ from the cavity and suppressing vibration-vibration entanglement. Similarly, when $M=N$, $G_2=0$, decoupling mode $B_2$. Maximal vibration-vibration entanglement is achieved when $G_1 \approx G_2$ (i.e., $M \approx \frac N2$), resulting in a balanced and symmetric coupling of both collective vibration modes to the cavity field, thus maximizing indirect coupling and entanglement between $B_1$ and $B_2$. This result aligns with theoretical predictions for entanglement optimization in hybrid quantum systems with balanced coupling strengths~\cite{Huang2024}.
	
	\subsection{Entanglement dependence on cavity detunings and OPA enhancement}\label{sec:detunings}
	
	\begin{figure}[htp!]
		\centering
		\includegraphics[width=9.1cm]{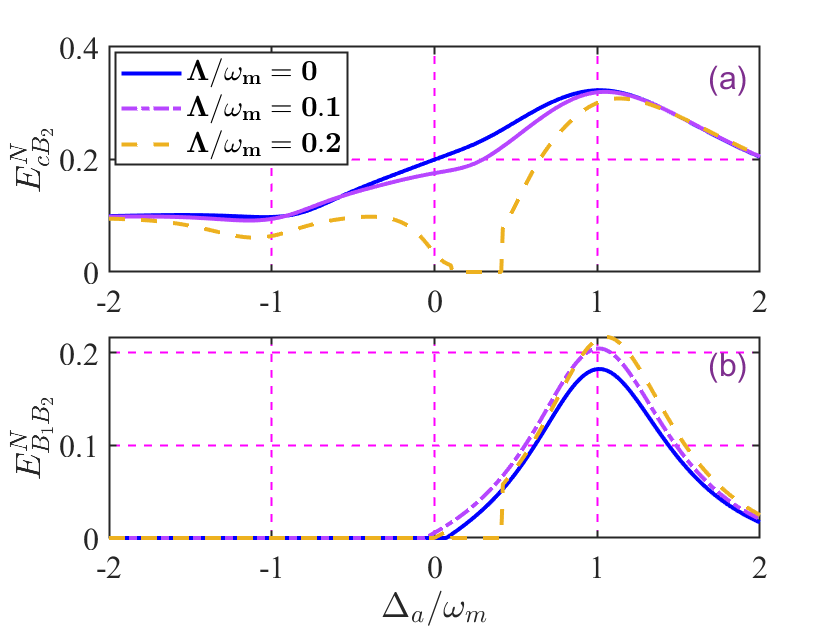}	
		\caption{(a) optical-vibration entanglement $E^N_{cB_2}$ and (b) Vibration-vibration entanglement $E_{B_1B_2}^N$ as a function of the normalized cavity detuning $\Delta_a/\omega_m$ (for Cavity 1). Entanglement is quantified by Logarithmic Negativity. Parameters are the same as for \Cref{fig:fig2} except for the driving amplitude $\frac{\mathcal{E}}{\omega_m}=16$. Optimal entanglement is achieved near resonance ($\Delta_a/\omega_m \approx 1$).}
		\label{fig:fig6}
	\end{figure}
	
	To investigate the impact of cavity detuning on entanglement, we analyze bipartite entanglement as a function of normalized cavity detuning $\Delta_a/\omega_m$ (for Cavity 1) and $\tilde{\Delta}_c/\omega_m$ (for Cavity 2).
	
	\Cref{fig:fig6} presents (a) optical-vibration entanglement $E^N_{cB_2}$ and (b) vibration-vibration entanglement $E_{B_1B_2}^N$ as a function of normalized cavity detuning $\Delta_a/\omega_m$ for different OPA nonlinear gains $\frac{\Lambda}{\omega_m}$. The results indicate that the optimal detuning for achieving maximum entanglement values is located near the resonance, that is, $\Delta_a/\omega_m \approx 1$. This can be physically understood as follows: resonant driving of Cavity 1 enhances the intracavity field strength and optomechanical coupling, leading to stronger entanglement. Furthermore, \Cref{fig:fig6}(a) shows that a larger detuning range supports optical-vibration entanglement compared to vibration-vibration entanglement (\Cref{fig:fig6}(b)). This broader detuning tolerance for optical-vibration entanglement may be advantageous in experimental settings, offering greater flexibility in tuning system parameters and maintaining entanglement under realistic experimental imperfections. However, the generation of an enhanced vibration-vibration entanglement requires a non-zero OPA gain ($\frac{\Lambda}{\omega_m} = 0.2$ in \Cref{fig:fig6}). The introduction of OPA, while enhancing vibration-vibration entanglement (see \Cref{fig:fig5}(a), \Cref{fig:fig6}(a)), is again observed to be detrimental to optical-vibration entanglement (see \Cref{fig:fig5}(b), \Cref{fig:fig6}(b)), reinforcing the counterintuitive trade-off we have identified.
	
	\begin{figure}[htp!]
		\centering
		\includegraphics[width=9.1cm]{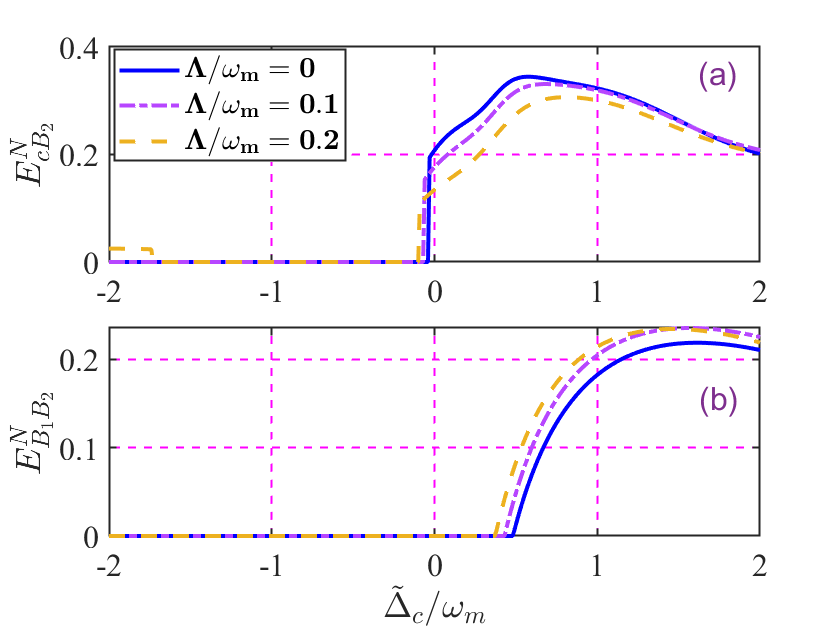}
		\caption{(a) optical-vibration entanglement $E^N_{cB_2}$ and (b) vibration-vibration entanglement $E_{B_1B_2}^N$ as a function of the normalized cavity detuning $\tilde{\Delta}_c/\omega_m$ (for Cavity 2). Entanglement is quantified by Logarithmic Negativity. Parameters are the same as for \Cref{fig:fig2} except for the driving amplitude $\frac{\mathcal{E}}{\omega_m}=16$. Optimal detunings for peak entanglement are shifted, with vibration-vibration entanglement peaking at larger detuning ($\tilde{\Delta}_c/\omega_m \approx 1.5$) compared to optical-vibration entanglement ($\tilde{\Delta}_c/\omega_m \approx 0.5$).}
		\label{fig:fig7}
	\end{figure}
	
	\Cref{fig:fig7} plots both (a) optical-vibration entanglement $E^N_{cB_2}$ and (b) vibration-vibration entanglement $E_{B_1B_2}^N$ as a function of normalized cavity detuning $\tilde{\Delta}_c/\omega_m$ of Cavity 2. The optimal detuning for the peak $E^N_{cB_2}$ is found to be around $\tilde{\Delta}_c/\omega_m \approx 0.5$, while for $E_{B_1B_2}^N$, the peak entanglement occurs with a larger detuning of $\tilde{\Delta}_c/\omega_m \approx 1.5$. The shift in optimal detuning for vibration-vibration entanglement compared to optical-vibration entanglement suggests that different detuning conditions may be required to optimize entanglement within different subsystems of the McOM system. This could be exploited in experiments to selectively enhance or suppress different types of entanglement by adjusting the driving frequencies.
	
	\subsection{Robustness to thermal noise}\label{sec:thermal}
	
\begin{figure}[htp!]
		\centering
		\includegraphics[width=9.1cm]{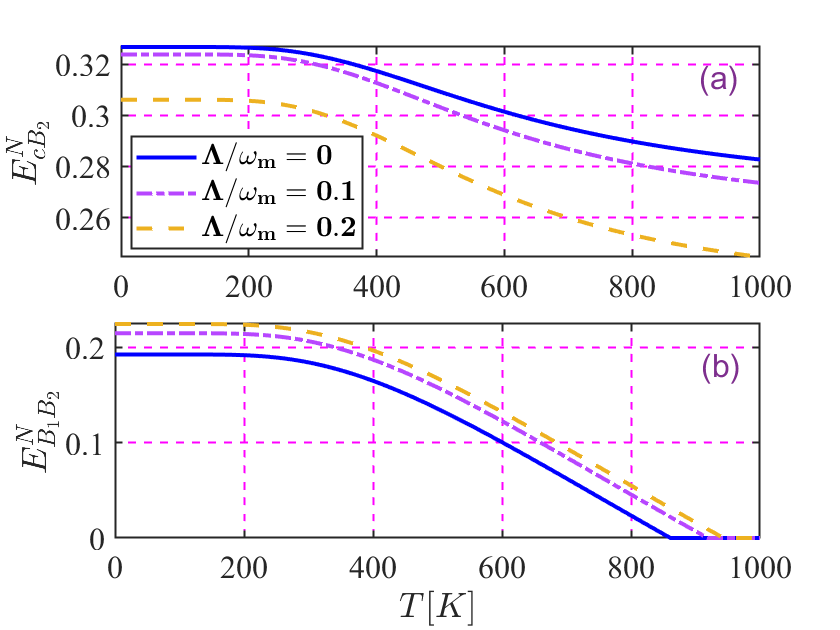}	
		\caption{(a) optical-vibration entanglement $E^N_{cB_2}$ and  (b) vibration-vibration entanglement $E_{B_1B_2}^N$ as a function of temperature $T$. Entanglement is quantified by Logarithmic Negativity. Parameters are the same as for \Cref{fig:fig2} except for the driving amplitude $\frac{\mathcal{E}}{\omega_m}=16$. Vibration-vibration entanglement exhibits remarkable robustness, persisting to near-room temperatures and enhanced by the presence of the OPA.}
		\label{fig:fig8}
	\end{figure}

	Temperature is a critical factor for entanglement in practical quantum systems, as thermal noise can rapidly degrade quantum correlations. \Cref{fig:fig8} assesses the robustness of the entanglement in our McOM system to temperature variations, plotting both the optical-vibration and vibration-vibration entanglement as a function of temperature $T$, in the presence and absence of OPA.
	
	As expected, \Cref{fig:fig8} shows a general decrease in both types of entanglement with increasing temperature, due to the increasing population of thermal phonons and associated decoherence. However, vibration-vibration entanglement exhibits remarkable resilience, persisting up to temperatures approaching $T \approx \SI{e3}{\kelvin}$. This high-temperature robustness is a significant advantage of McOM systems, stemming from the high vibrational frequencies of molecules, which reduce thermal phonon occupation even at room temperature. The optical vibration entanglement, although also robust, degrades more rapidly with temperature and is further weakened by the presence of OPA (\Cref{fig:fig8}(a)). In contrast, the OPA increases the robustness of the vibration-vibration coupling against thermal fluctuations (\Cref{fig:fig8}(b)), further highlighting the differential impact of the OPA on different coupling subsystems. The ability of vibration-vibration entanglement to endure at elevated temperatures, particularly with OPA enhancement, makes our proposed McOM scheme highly promising for practical applications in quantum computing and communication, where high operational temperatures are often unavoidable.
	
	\subsection{Phase control of entanglement via OPA}\label{sec:phase}
	
\begin{figure}[htp!]
		\centering
		\includegraphics[width=9.1cm]{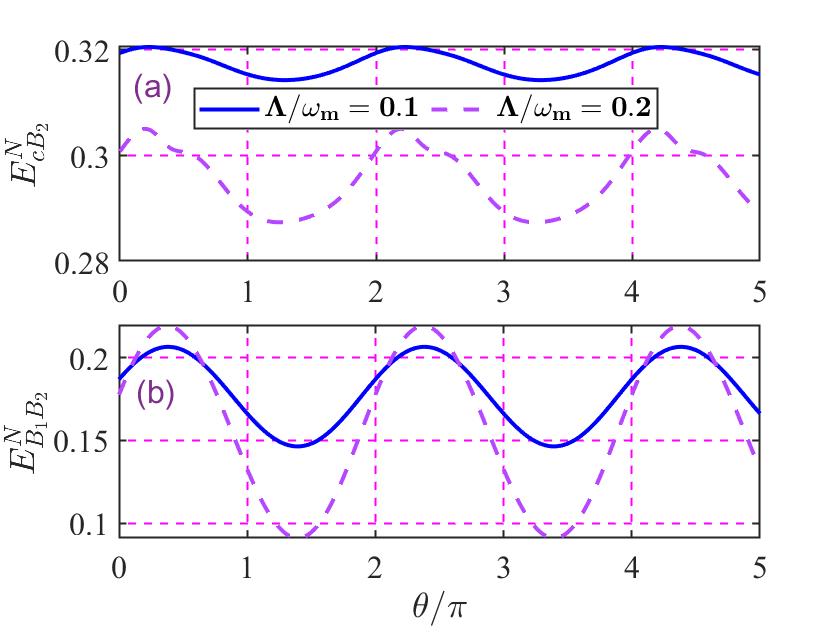}
		\caption{(a) optical-vibration entanglement $E^N_{cB_2}$ versus parametric amplifier phase $\theta$, and (b) vibration-vibration entanglement $E_{B_1B_2}^N$ versus parametric amplifier phase $\theta$. Entanglement is quantified by Logarithmic Negativity. Parameters are the same as for \Cref{fig:fig2} except for the driving amplitude $\frac{\mathcal{E}}{\omega_m}=16$. Entanglement exhibits periodic dependence on $\theta$, maximized at specific phase values.}
		\label{fig:fig9}
	\end{figure}

	The phase $\theta$ of the optical field driving the OPA is another important control parameter in our system. \Cref{fig:fig9} explores the dependence of bipartite entanglement on the parametric amplifier phase $\theta$.
	
	\Cref{fig:fig9} reveals a strong dependence of optical-vibration and vibration-vibration entanglement on the OPA phase $\theta$. In particular, entanglement is periodically maximized with respect to $\theta$, showing peak values around $\theta \approx (2n + \frac12)\pi$ (where $n$ is an integer) and minima around $\theta \approx (2n + \frac{3}{2})\pi$. This periodic dependence arises from the phase sensitive nature of the OPA process, where the phase $\theta$ directly influences the parametric amplification and squeezing of quantum fluctuations, thus modulating the generated entanglement. The existence of optimal phase values for the enhancement of the entanglement suggests that precise phase control of the OPA drive is essential to maximize entanglement in experimental implementations. While non-zero entanglement can be achieved without OPA ($\Lambda=0$), as shown in \Crefrange{fig:fig3}{fig:fig8}, the presence of OPA with optimal phase significantly enhances the achievable entanglement levels, particularly for vibration-vibration entanglement. This underscores the critical role of OPA in \textit{boosting} entanglement in this system.
	
	\subsection{Impact of nonreciprocal cavity coupling on bipartite entanglement}\label{sec:results_nonreciprocal}
	
\begin{figure}[htp!]
		\centering
		\includegraphics[width=9.1cm]{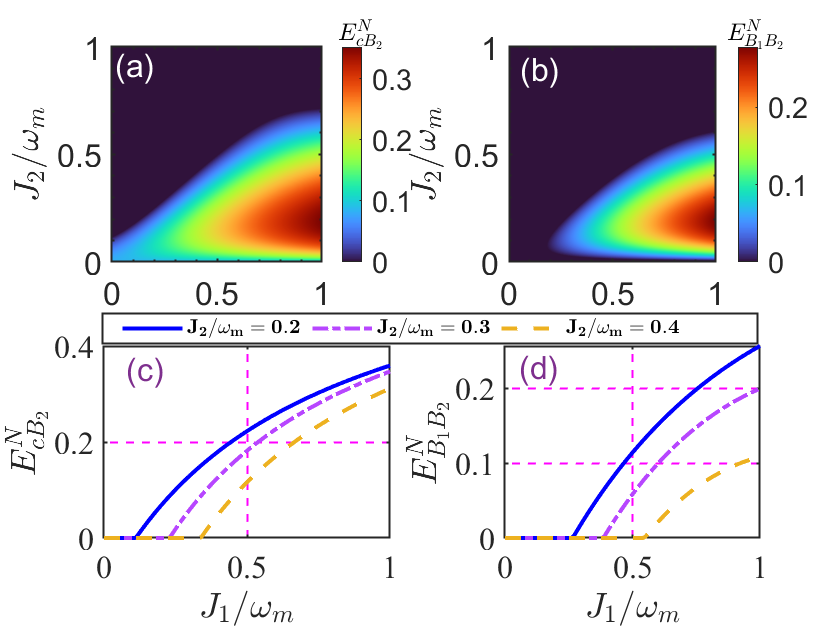}	
		\caption{(a) Contour plot of optical-vibration entanglement $E^N_{cB_2}$ versus nonreciprocal coupling $J_2/\omega_m$ and $J_1/\omega_m$,  (b) Contour plot of vibration-vibration entanglement $E^N_{B_1B_2}$ versus  nonreciprocal coupling $J_2/\omega_m$ and $J_1/\omega_m$, (c) photon-vibration entanglement $E^N_{cB_2}$ versus nonreciprocal coupling $J_1/\omega_m$, (d) vibration-vibration entanglement $E^N_{B_1B_2}$ versus nonreciprocal coupling $J_1/\omega_m$. Entanglement is quantified by Logarithmic Negativity. Parameters are the same as for \Cref{fig:fig2} except for the driving amplitude $\frac{\mathcal{E}}{\omega_m}=16$ and nonlinear gain $\frac{\Lambda}{\omega_m}=0.2$. Nonreciprocal coupling ($J_1 \neq J_2$) is essential for significant entanglement, with optimal entanglement achieved when $J_1 > J_2$.}
		\label{fig:fig10}
	\end{figure}

	Finally, we investigate the role of nonreciprocal coupling between the two cavities, a key feature of our proposed McOM architecture. \Cref{fig:fig10} presents contour plots of bipartite entanglement as a function of coupling strengths $J_1/\omega_m$ (coupling from cavity 1 to cavity 2) and $J_2/\omega_m$ (coupling from Cavity 2 to Cavity 1).
	
	\Cref{fig:fig10} explores the impact of non-reciprocal coupling strengths, $J_1$ and $J_2$, on bipartite entanglement. The contour plots of cavity vibration entanglement $E^N_{cB_2}$ (\Cref{fig:fig10} (a)) and vibration-vibration entanglement $E_{B_1B_2}^N$ (\Cref{fig:fig10}(b)) as functions of $J_1/\omega_m$ and $J_2/\omega_m$ reveal that both types of entanglement are significantly suppressed when cavities are reciprocally coupled (that is, $J_1 \approx J_2$). This suppression of entanglement under reciprocal coupling underscores the importance of non-reciprocity for entanglement generation in our double-cavity McOM system. Physically, non-reciprocal coupling, establishing a directional flow of quantum information, prevents destructive interference effects that can arise in reciprocally coupled systems and degrade entanglement.
	
	The optimal non-reciprocal coupling strengths for maximizing entanglement are located around $J_1/\omega_m \approx 1$ and $J_2/\omega_m \approx 0.3$. This asymmetry in optimal coupling strengths suggests that a stronger coupling in the forward direction ($J_1$, from Cavity 1 to Cavity 2) and a weaker coupling in the backward direction ($J_2$, from Cavity 2 to Cavity 1) are conducive to entanglement generation. This can be further visualized in Figures~\ref{fig:fig10}(c) and (d), which plot $E^N_{cB_2}$ and $E_{B_1B_2}^N$ as functions of $J_1/\omega_m$ for different values of $J_2/\omega_m$. These plots clearly demonstrate that the entanglement is significantly improved when $J_1 > J_2$, reinforcing the necessity of non-reciprocal coupling for the efficient generation of entanglement in our system. The optimal asymmetry in coupling strengths could be engineered in practice by carefully designing the waveguide coupling section between the Fabry-Pérot and plasmonic cavities to achieve the desired directional interaction.
	
	\subsection{Discussion and implications for quantum information processing}\label{sec:discussion}
	
	Our analysis reveals several key findings with significant implications for quantum information processing using molecular cavity optomechanical systems. (i) First, we demonstrate that the integration of an OPA within a McOM system provides a powerful mechanism to enhance vibration-vibration entanglement, a critical resource for quantum memories and transducers. This enhancement opens new avenues for engineering robust entanglement sources for quantum networks and distributed quantum computing architectures. However, this enhancement is not without trade-offs. Our results reveal a counterintuitive suppression of optical-vibration entanglement in the presence of the OPA. This trade-off highlights the complex interplay of non-linear optics and optomechanics in McOM systems and underscores the need for careful system design tailored to specific entanglement objectives. For applications prioritizing cavity-based readout or transduction, alternative entanglement enhancement strategies may be warranted. (ii) Second, our finding that vibration-vibration entanglement is maximized under a symmetric population of molecular collective modes provides a clear and actionable guideline for experimental optimization. Precise control over the molecular distribution within McOM devices, which can be achieved through microfluidic control or patterned de-position, could be leveraged to maximize the efficiency of entanglement generation. (iii) Thirdly, and perhaps most significantly for practical quantum technologies, we demonstrate the remarkable robustness of vibration-vibration entanglement in our OPA-enhanced McOM system to thermal noise, with entanglement persisting at temperatures approaching \SI{e3}{\kelvin}. This resilience, exceeding that of conventional COM systems, is a major advantage for McOM platforms and positions our scheme as particularly promising for real-world quantum information processing applications operating at ambient temperatures. (iv) Finally, the critical role of nonreciprocal cavity coupling in our system design highlights the importance of carefully engineering inter-cavity interactions for entanglement generation in hybrid quantum systems. The demonstrated sensitivity of entanglement to the asymmetry in coupling strengths $J_1$ and $J_2$ offers a valuable control knob to tailor entanglement properties.
	
	These findings collectively provide a valuable theoretical roadmap for exploiting OPA-enhanced McOM systems in quantum information processing. The ability to enhance and control vibration-vibration entanglement, coupled with thermal robustness and design flexibility offered by OPA and cavity parameters, positions this architecture as a promising candidate for realizing robust and scalable quantum devices, leveraging the inherent scalability of molecular self-assembly and the potential for integration with existing photonic circuits. 
	
	Although our theoretical model provides a comprehensive framework for understanding entanglement in this system, it also has limitations.  Our model simplifies the molecular system by considering only two collective vibration modes and neglects potential complexities arising from anharmonicities or higher-order vibrational modes, as discussed in Ref.~\cite{Xiang2024}. Furthermore, we have assumed ideal cavity conditions and perfect nonreciprocal coupling, which may deviate from experimental realities. Future research could explore the impact of these factors on entanglement and investigate more complex McOM architectures with multiple cavities and multimode interactions. Experimentally, achieving the proposed scheme requires the fabrication of double-cavity McOM systems with integrated OPAs and tunable non-reciprocal coupling, which presents a significant but potentially surmountable technological challenge. However, recent advances in nanofabrication and integrated quantum photonics, as highlighted by ~\textcite{Roelli2024}, suggest that such experimental implementations may become feasible in the near future. Further theoretical work could also explore alternative nonlinear elements beyond OPAs, and investigate the potential of this system for generating multipartite entanglement and other advanced quantum states.
	
	\section{Conclusion}\label{sec:Concl}
	
	This work introduces and rigorously analyzes a novel theoretical framework for realizing enhanced quantum entanglement, a critical resource for advancing quantum information processing, within a double-cavity molecular optomechanical (McOM) system incorporating an optical parametric amplifier (OPA). Our findings unequivocally demonstrate that OPAs offer a potent tool for significantly boosting vibration-vibration entanglement ($E_{B_1B_2}^N$), a key resource for quantum memories and transducers, in McOM systems. This represents a critical step towards harnessing McOM systems for quantum memories and transducers in distributed quantum networks. Importantly, we reveal a counterintuitive trade-off: this enhancement of vibration-vibration entanglement is accompanied by a concomitant suppression of optical-vibration entanglement ($E_{cB_2}^N$), which is essential for quantum state transfer and readout, a key design consideration for McOM-based quantum information architectures. Furthermore, our analysis pinpoints the symmetric molecular collective mode population as the optimal condition for maximizing vibration-vibration entanglement, offering a clear and actionable route for experimental optimization of entanglement sources. Remarkably, we establish the inherent thermal robustness of vibration-vibration entanglement in our OPA-enhanced McOM system, which is capable of surviving at temperatures approaching \SI{e3}{\kelvin}. This resilience, significantly exceeding the thermal resilience of conventional COM systems, underscores the practical viability of our scheme for room-temperature quantum technologies.
	
	This work establishes a promising theoretical foundation for exploiting OPA-enhanced McOM systems as a robust and scalable platform for quantum information processing and quantum computing. The demonstrated enhancement and thermal resilience of vibration-vibration entanglement, coupled with the inherent tunability afforded by OPA gain, phase, and cavity coupling parameters, opens compelling avenues for realizing practical, high-performance quantum devices operating at ambient temperatures. Future research should prioritize experimental validation of these theoretical predictions, focusing on the fabrication of integrated McOM-OPA devices and the exploration of multipartite entanglement protocols for complex quantum information tasks. Furthermore, theoretical investigations into alternative nonlinear optical elements and the role of molecular anharmonicities will be essential for pushing the boundaries of entanglement generation and control in McOM systems, ultimately unlocking their full potential for revolutionizing quantum technologies. Our study thus paves the way for a new generation of quantum devices based on molecular cavity optomechanics, capable of addressing key challenges in quantum information processing and quantum computing.
	
	\section*{Acknowledgments}
	
	P.D. acknowledges the Iso-Lomso Fellowship at Stellenbosch Institute for Advanced Study (STIAS), Wallenberg Research Centre at Stellenbosch University, Stellenbosch 7600, South Africa, and The Institute for Advanced Study, Wissenschaftskolleg zu Berlin, Wallotstrasse 19, 14193 Berlin, Germany.  K.S. Nisar thanks the funding from Prince Sattam bin Abdulaziz University, Saudi Arabia project number (PSAU/2024/R/1445). The authors thank the Dean of Graduate Studies and Scientific Research at the University of Bisha for supporting this work through the Fast-Track Research Support Program.
	
	\textbf{Author Contributions:} E.K.B. conceptualized the work and carried out the simulations and analysis. C.T. conceptualized the work. P.D. and S.G.N.E.  participated in all the discussions and provided useful suggestions to the final version of the manuscript. A.-H. A.-A. and K.S.N. supervised the work. All authors participated equally in writing, discussions, editing, and review of the manuscript.
	
	\textbf{Competing Interests:} All authors declare no competing interests.
	
	\textbf{Data Availability:}
	Relevant data are included in the manuscript and supporting information. Supplement data are available upon reasonable request.
	
	\appendix
	\section{Derivation of the Hamiltonian}\label{sec:Ham}
	
	The Hamiltonian of the molecular optomechanical system in question reads ($\hbar=1$)
	\begin{equation} \label{eq:H_full}
	\begin{aligned}
	H(t) = & \omega_a a^\dagger a + \omega_c c^\dagger c + \sum_{j=1}^N \omega_m b_j^\dagger b_j - \sum_{j=1}^N g_m c^\dagger c (b_j^\dagger + b_j) +\\& J_1 a^\dagger c+ J_2 c^\dagger a  +i\Lambda \left( e^{i\theta} a^{\dagger 2} e^{-2i\omega_{01}t} - e^{-i\theta} a^2 e^{2i\omega_{01}t} \right)\\&+ i\mathcal{E}_a \left( a^\dagger e^{-i\omega_L t} - a e^{i\omega_L t} \right) + i\mathcal{E}_c \left( c^\dagger e^{-i\omega_L t} - c e^{i\omega_L t} \right) .
	\end{aligned}
	\end{equation}
	To move into a rotating frame at frequency $\omega_L$, we define the unitary transformation:
	\begin{equation}
	U(t) = \exp[i \omega_L t (a^\dagger a + c^\dagger c)] .
	\end{equation}
	The transformed Hamiltonian is given by:
	\begin{equation}
	\tilde{H}(t) = U(t) H(t) U^\dagger(t) - i U(t) \frac{d}{dt} U^\dagger(t).
	\end{equation}
	We compute the derivative term:
	\begin{equation}
	-i U(t) \frac{d}{dt} U^\dagger(t) = -\omega_L (a^\dagger a + c^\dagger c),
	\end{equation}
	thus, the rotating-frame Hamiltonian becomes:
	\begin{equation}
	\tilde{H}(t) = U H U^\dagger - \omega_L (a^\dagger a + c^\dagger c).
	\end{equation}
	The transformed terms are the following:
	\begin{itemize}
		\item \textbf{Free Hamiltonian.}
		\begin{equation}
		\omega_a a^\dagger a + \omega_c c^\dagger c - \omega_L (a^\dagger a + c^\dagger c)
		= \Delta_a a^\dagger a + \Delta_c c^\dagger c,
		\end{equation}
		where
		$\Delta_a = \omega_a - \omega_L, $ and $ \Delta_c = \omega_c - \omega_L$.
		
		\item  \textbf{Mechanical terms.} These remain unchanged:
		\begin{equation}
		\sum_{j=1}^N \omega_m b_j^\dagger b_j - \sum_{j=1}^N g_m c^\dagger c (b_j^\dagger + b_j).
		\end{equation}
		\item  \textbf{Mode coupling terms} These remain unchanged:
		\begin{equation}
		J_1 a^\dagger c + J_2 c^\dagger a.
		\end{equation}
		
		\item \textbf{Squeezing term.} Transforming under $a \to a e^{-i\omega_L t}$, we obtain the following:
		\begin{equation}
		a^{\dagger 2} e^{-2i\omega_{01}t} \to a^{\dagger 2} e^{2i(\omega_L - \omega_{01})t}.
		\end{equation}
		Hence the squeezing term becomes:
		\begin{equation}
		i\Lambda \left( e^{i\theta} a^{\dagger 2} e^{2i(\omega_L - \omega_{01})t} - e^{-i\theta} a^2 e^{-2i(\omega_L - \omega_{01})t} \right).
		\end{equation}
	\end{itemize}
	
	Under the \textbf{Rotating Wave Approximation (RWA)}, if $\omega_L \approx \omega_{01}$, the fast oscillating terms are dropped:
	\begin{equation}
	\approx i\Lambda \left( e^{i\theta} a^{\dagger 2} - e^{-i\theta} a^2 \right).
	\end{equation}
	
	The drive terms are transformed as:
	\begin{align}
	i\mathcal{E}_a \left( a^\dagger e^{-i\omega_L t} - a e^{i\omega_L t} \right) &\to i\mathcal{E}_a (a^\dagger - a), \\
	i\mathcal{E}_c \left( c^\dagger e^{-i\omega_L t} - c e^{i\omega_L t} \right) &\to i\mathcal{E}_c (c^\dagger - c).
	\end{align}
	
	Combining all terms, the effective time-independent Hamiltonian in the rotating frame under RWA is:
	\begin{equation} \label{eq:H_rot}
	\begin{aligned}
	\tilde{H}_{\mathrm{new}} = &\Delta_a a^\dagger a + \Delta_c c^\dagger c + \sum_{j=1}^N \omega_m b_j^\dagger b_j\\
	& - \sum_{j=1}^N g_m c^\dagger c (b_j^\dagger + b_j) + J_1 a^\dagger c + J_2 c^\dagger a  \\
	& + i\Lambda \left( e^{i\theta} a^{\dagger 2} - e^{-i\theta} a^2 \right) + i\mathcal{E}_a (a^\dagger - a)\\
	&+ i\mathcal{E}_c (c^\dagger - c).
	\end{aligned}
	\end{equation}
	
	\bibliography{RefMolecularOPA}

\begin{thebibliography}{38}%
\makeatletter
\providecommand \@ifxundefined [1]{%
 \@ifx{#1\undefined}
}%
\providecommand \@ifnum [1]{%
 \ifnum #1\expandafter \@firstoftwo
 \else \expandafter \@secondoftwo
 \fi
}%
\providecommand \@ifx [1]{%
 \ifx #1\expandafter \@firstoftwo
 \else \expandafter \@secondoftwo
 \fi
}%
\providecommand \natexlab [1]{#1}%
\providecommand \enquote  [1]{``#1''}%
\providecommand \bibnamefont  [1]{#1}%
\providecommand \bibfnamefont [1]{#1}%
\providecommand \citenamefont [1]{#1}%
\providecommand \href@noop [0]{\@secondoftwo}%
\providecommand \href [0]{\begingroup \@sanitize@url \@href}%
\providecommand \@href[1]{\@@startlink{#1}\@@href}%
\providecommand \@@href[1]{\endgroup#1\@@endlink}%
\providecommand \@sanitize@url [0]{\catcode `\\12\catcode `\$12\catcode
  `\&12\catcode `\#12\catcode `\^12\catcode `\_12\catcode `\%12\relax}%
\providecommand \@@startlink[1]{}%
\providecommand \@@endlink[0]{}%
\providecommand \url  [0]{\begingroup\@sanitize@url \@url }%
\providecommand \@url [1]{\endgroup\@href {#1}{\urlprefix }}%
\providecommand \urlprefix  [0]{URL }%
\providecommand \Eprint [0]{\href }%
\providecommand \doibase [0]{https://doi.org/}%
\providecommand \selectlanguage [0]{\@gobble}%
\providecommand \bibinfo  [0]{\@secondoftwo}%
\providecommand \bibfield  [0]{\@secondoftwo}%
\providecommand \translation [1]{[#1]}%
\providecommand \BibitemOpen [0]{}%
\providecommand \bibitemStop [0]{}%
\providecommand \bibitemNoStop [0]{.\EOS\space}%
\providecommand \EOS [0]{\spacefactor3000\relax}%
\providecommand \BibitemShut  [1]{\csname bibitem#1\endcsname}%
\let\auto@bib@innerbib\@empty
\bibitem [{\citenamefont {Lai}\ \emph {et~al.}(2022)\citenamefont {Lai},
  \citenamefont {Liao}, \citenamefont {Miranowicz},\ and\ \citenamefont
  {Nori}}]{Lai2022}%
  \BibitemOpen
  \bibfield  {author} {\bibinfo {author} {\bibfnamefont {D.-G.}\ \bibnamefont
  {Lai}}, \bibinfo {author} {\bibfnamefont {J.-Q.}\ \bibnamefont {Liao}},
  \bibinfo {author} {\bibfnamefont {A.}~\bibnamefont {Miranowicz}},\ and\
  \bibinfo {author} {\bibfnamefont {F.}~\bibnamefont {Nori}},\ }\bibfield
  {title} {\bibinfo {title} {Noise-tolerant optomechanical entanglement via
  synthetic magnetism},\ }\href
  {https://doi.org/10.1103/PhysRevLett.129.063602} {\bibfield  {journal}
  {\bibinfo  {journal} {Physical Review Letters}\ }\textbf {\bibinfo {volume}
  {129}},\ \bibinfo {pages} {063602} (\bibinfo {year} {2022})}\BibitemShut
  {NoStop}%
\bibitem [{\citenamefont {Tchodimou}\ \emph {et~al.}(2017)\citenamefont
  {Tchodimou}, \citenamefont {Djorwe},\ and\ \citenamefont
  {Nana~Engo}}]{Tchodimou2017}%
  \BibitemOpen
  \bibfield  {author} {\bibinfo {author} {\bibfnamefont {C.}~\bibnamefont
  {Tchodimou}}, \bibinfo {author} {\bibfnamefont {P.}~\bibnamefont {Djorwe}},\
  and\ \bibinfo {author} {\bibfnamefont {S.~G.}\ \bibnamefont {Nana~Engo}},\
  }\bibfield  {title} {\bibinfo {title} {Distant entanglement enhanced in
  $\mathcal{PT}$-symmetric optomechanics},\ }\href
  {https://doi.org/10.1103/PhysRevA.96.033856} {\bibfield  {journal} {\bibinfo
  {journal} {Phys. Rev. A}\ }\textbf {\bibinfo {volume} {96}},\ \bibinfo
  {pages} {033856} (\bibinfo {year} {2017})}\BibitemShut {NoStop}%
\bibitem [{\citenamefont {Araya-Sossa}\ and\ \citenamefont
  {Orszag}(2023)}]{Araya2023}%
  \BibitemOpen
  \bibfield  {author} {\bibinfo {author} {\bibfnamefont {K.}~\bibnamefont
  {Araya-Sossa}}\ and\ \bibinfo {author} {\bibfnamefont {M.}~\bibnamefont
  {Orszag}},\ }\bibfield  {title} {\bibinfo {title} {Generation of entanglement
  via squeezing on a tripartite-optomechanical system},\ }\href
  {https://doi.org/10.1103/PhysRevA.108.012432} {\bibfield  {journal} {\bibinfo
   {journal} {Physical Review A}\ }\textbf {\bibinfo {volume} {108}},\ \bibinfo
  {pages} {012432} (\bibinfo {year} {2023})}\BibitemShut {NoStop}%
\bibitem [{\citenamefont {Agasti}\ and\ \citenamefont
  {Djorwé}(2024)}]{Agasti2024}%
  \BibitemOpen
  \bibfield  {author} {\bibinfo {author} {\bibfnamefont {S.}~\bibnamefont
  {Agasti}}\ and\ \bibinfo {author} {\bibfnamefont {P.}~\bibnamefont
  {Djorwé}},\ }\bibfield  {title} {\bibinfo {title} {Bistability-assisted
  mechanical squeezing and entanglement},\ }\href
  {https://doi.org/10.1088/1402-4896/ad6eca} {\bibfield  {journal} {\bibinfo
  {journal} {Physica Scripta}\ }\textbf {\bibinfo {volume} {99}},\ \bibinfo
  {pages} {095129} (\bibinfo {year} {2024})}\BibitemShut {NoStop}%
\bibitem [{\citenamefont {Ye}\ \emph {et~al.}(2025)\citenamefont {Ye},
  \citenamefont {Arenz}, \citenamefont {Lukens},\ and\ \citenamefont
  {Lai}}]{Ye2025}%
  \BibitemOpen
  \bibfield  {author} {\bibinfo {author} {\bibfnamefont {L.-L.}\ \bibnamefont
  {Ye}}, \bibinfo {author} {\bibfnamefont {C.}~\bibnamefont {Arenz}}, \bibinfo
  {author} {\bibfnamefont {J.~M.}\ \bibnamefont {Lukens}},\ and\ \bibinfo
  {author} {\bibfnamefont {Y.-C.}\ \bibnamefont {Lai}},\ }\bibfield  {title}
  {\bibinfo {title} {Entanglement engineering of optomechanical systems by
  reinforcement learning},\ }\bibfield  {journal} {\bibinfo  {journal} {APL
  Machine Learning}\ }\textbf {\bibinfo {volume} {3}},\ \href
  {https://doi.org/10.1063/5.0233470} {10.1063/5.0233470} (\bibinfo {year}
  {2025})\BibitemShut {NoStop}%
\bibitem [{\citenamefont {Chen}\ \emph {et~al.}(2025)\citenamefont {Chen},
  \citenamefont {Luo},\ and\ \citenamefont {Yu}}]{Chen2025}%
  \BibitemOpen
  \bibfield  {author} {\bibinfo {author} {\bibfnamefont {P.}~\bibnamefont
  {Chen}}, \bibinfo {author} {\bibfnamefont {D.-W.}\ \bibnamefont {Luo}},\ and\
  \bibinfo {author} {\bibfnamefont {T.}~\bibnamefont {Yu}},\ }\bibfield
  {title} {\bibinfo {title} {Optimal entanglement generation in optomechanical
  systems via krotov control of covariance matrix dynamics},\ }\href
  {https://doi.org/10.1103/PhysRevResearch.7.013161} {\bibfield  {journal}
  {\bibinfo  {journal} {Physical Review Research}\ }\textbf {\bibinfo {volume}
  {7}},\ \bibinfo {pages} {013161} (\bibinfo {year} {2025})}\BibitemShut
  {NoStop}%
\bibitem [{\citenamefont {Kotler}\ \emph {et~al.}(2021)\citenamefont {Kotler},
  \citenamefont {Peterson}, \citenamefont {Shojaee}, \citenamefont {Lecocq},
  \citenamefont {Cicak}, \citenamefont {Kwiatkowski}, \citenamefont {Geller},
  \citenamefont {Glancy}, \citenamefont {Knill}, \citenamefont {Simmonds},
  \citenamefont {Aumentado},\ and\ \citenamefont {Teufel}}]{Kotler2021}%
  \BibitemOpen
  \bibfield  {author} {\bibinfo {author} {\bibfnamefont {S.}~\bibnamefont
  {Kotler}}, \bibinfo {author} {\bibfnamefont {G.~A.}\ \bibnamefont
  {Peterson}}, \bibinfo {author} {\bibfnamefont {E.}~\bibnamefont {Shojaee}},
  \bibinfo {author} {\bibfnamefont {F.}~\bibnamefont {Lecocq}}, \bibinfo
  {author} {\bibfnamefont {K.}~\bibnamefont {Cicak}}, \bibinfo {author}
  {\bibfnamefont {A.}~\bibnamefont {Kwiatkowski}}, \bibinfo {author}
  {\bibfnamefont {S.}~\bibnamefont {Geller}}, \bibinfo {author} {\bibfnamefont
  {S.}~\bibnamefont {Glancy}}, \bibinfo {author} {\bibfnamefont
  {E.}~\bibnamefont {Knill}}, \bibinfo {author} {\bibfnamefont {R.~W.}\
  \bibnamefont {Simmonds}}, \bibinfo {author} {\bibfnamefont {J.}~\bibnamefont
  {Aumentado}},\ and\ \bibinfo {author} {\bibfnamefont {J.~D.}\ \bibnamefont
  {Teufel}},\ }\bibfield  {title} {\bibinfo {title} {Direct observation of
  deterministic macroscopic entanglement},\ }\href
  {https://doi.org/10.1126/science.abf2998} {\bibfield  {journal} {\bibinfo
  {journal} {Science}\ }\textbf {\bibinfo {volume} {372}},\ \bibinfo {pages}
  {622} (\bibinfo {year} {2021})}\BibitemShut {NoStop}%
\bibitem [{\citenamefont {Clarke}\ \emph {et~al.}(2020)\citenamefont {Clarke},
  \citenamefont {Sahium}, \citenamefont {Khosla}, \citenamefont {Pikovski},
  \citenamefont {Kim},\ and\ \citenamefont {Vanner}}]{Clarke2020}%
  \BibitemOpen
  \bibfield  {author} {\bibinfo {author} {\bibfnamefont {J.}~\bibnamefont
  {Clarke}}, \bibinfo {author} {\bibfnamefont {P.}~\bibnamefont {Sahium}},
  \bibinfo {author} {\bibfnamefont {K.~E.}\ \bibnamefont {Khosla}}, \bibinfo
  {author} {\bibfnamefont {I.}~\bibnamefont {Pikovski}}, \bibinfo {author}
  {\bibfnamefont {M.~S.}\ \bibnamefont {Kim}},\ and\ \bibinfo {author}
  {\bibfnamefont {M.~R.}\ \bibnamefont {Vanner}},\ }\bibfield  {title}
  {\bibinfo {title} {Generating mechanical and optomechanical entanglement via
  pulsed interaction and measurement},\ }\href
  {https://doi.org/10.1088/1367-2630/ab7ddd} {\bibfield  {journal} {\bibinfo
  {journal} {New Journal of Physics}\ }\textbf {\bibinfo {volume} {22}},\
  \bibinfo {pages} {063001} (\bibinfo {year} {2020})}\BibitemShut {NoStop}%
\bibitem [{\citenamefont {Rips}\ and\ \citenamefont
  {Hartmann}(2013)}]{Rips2013}%
  \BibitemOpen
  \bibfield  {author} {\bibinfo {author} {\bibfnamefont {S.}~\bibnamefont
  {Rips}}\ and\ \bibinfo {author} {\bibfnamefont {M.~J.}\ \bibnamefont
  {Hartmann}},\ }\bibfield  {title} {\bibinfo {title} {Quantum information
  processing with nanomechanical qubits},\ }\href
  {https://doi.org/10.1103/PhysRevLett.110.120503} {\bibfield  {journal}
  {\bibinfo  {journal} {Phys. Rev. Lett.}\ }\textbf {\bibinfo {volume} {110}},\
  \bibinfo {pages} {120503} (\bibinfo {year} {2013})}\BibitemShut {NoStop}%
\bibitem [{\citenamefont {Blais}\ \emph {et~al.}(2020)\citenamefont {Blais},
  \citenamefont {Girvin},\ and\ \citenamefont {Oliver}}]{Blais2020}%
  \BibitemOpen
  \bibfield  {author} {\bibinfo {author} {\bibfnamefont {A.}~\bibnamefont
  {Blais}}, \bibinfo {author} {\bibfnamefont {S.~M.}\ \bibnamefont {Girvin}},\
  and\ \bibinfo {author} {\bibfnamefont {W.~D.}\ \bibnamefont {Oliver}},\
  }\bibfield  {title} {\bibinfo {title} {Quantum information processing and
  quantum optics with circuit quantum electrodynamics},\ }\href
  {https://doi.org/10.1038/s41567-020-0806-z} {\bibfield  {journal} {\bibinfo
  {journal} {Nature Physics}\ }\textbf {\bibinfo {volume} {16}},\ \bibinfo
  {pages} {247} (\bibinfo {year} {2020})}\BibitemShut {NoStop}%
\bibitem [{\citenamefont {Xia}\ \emph {et~al.}(2023)\citenamefont {Xia},
  \citenamefont {Agrawal}, \citenamefont {Pluchar}, \citenamefont {Brady},
  \citenamefont {Liu}, \citenamefont {Zhuang}, \citenamefont {Wilson},\ and\
  \citenamefont {Zhang}}]{Xia2023}%
  \BibitemOpen
  \bibfield  {author} {\bibinfo {author} {\bibfnamefont {Y.}~\bibnamefont
  {Xia}}, \bibinfo {author} {\bibfnamefont {A.~R.}\ \bibnamefont {Agrawal}},
  \bibinfo {author} {\bibfnamefont {C.~M.}\ \bibnamefont {Pluchar}}, \bibinfo
  {author} {\bibfnamefont {A.~J.}\ \bibnamefont {Brady}}, \bibinfo {author}
  {\bibfnamefont {Z.}~\bibnamefont {Liu}}, \bibinfo {author} {\bibfnamefont
  {Q.}~\bibnamefont {Zhuang}}, \bibinfo {author} {\bibfnamefont {D.~J.}\
  \bibnamefont {Wilson}},\ and\ \bibinfo {author} {\bibfnamefont
  {Z.}~\bibnamefont {Zhang}},\ }\bibfield  {title} {\bibinfo {title}
  {Entanglement-enhanced optomechanical sensing},\ }\href
  {https://doi.org/10.1038/s41566-023-01178-0} {\bibfield  {journal} {\bibinfo
  {journal} {Nature Photonics}\ }\textbf {\bibinfo {volume} {17}},\ \bibinfo
  {pages} {470} (\bibinfo {year} {2023})}\BibitemShut {NoStop}%
\bibitem [{\citenamefont {Brady}\ \emph {et~al.}(2023)\citenamefont {Brady},
  \citenamefont {Chen}, \citenamefont {Xia}, \citenamefont {Manley},
  \citenamefont {Dey~Chowdhury}, \citenamefont {Xiao}, \citenamefont {Liu},
  \citenamefont {Harnik}, \citenamefont {Wilson}, \citenamefont {Zhang},\ and\
  \citenamefont {Zhuang}}]{Brady2023}%
  \BibitemOpen
  \bibfield  {author} {\bibinfo {author} {\bibfnamefont {A.~J.}\ \bibnamefont
  {Brady}}, \bibinfo {author} {\bibfnamefont {X.}~\bibnamefont {Chen}},
  \bibinfo {author} {\bibfnamefont {Y.}~\bibnamefont {Xia}}, \bibinfo {author}
  {\bibfnamefont {J.}~\bibnamefont {Manley}}, \bibinfo {author} {\bibfnamefont
  {M.}~\bibnamefont {Dey~Chowdhury}}, \bibinfo {author} {\bibfnamefont
  {K.}~\bibnamefont {Xiao}}, \bibinfo {author} {\bibfnamefont {Z.}~\bibnamefont
  {Liu}}, \bibinfo {author} {\bibfnamefont {R.}~\bibnamefont {Harnik}},
  \bibinfo {author} {\bibfnamefont {D.~J.}\ \bibnamefont {Wilson}}, \bibinfo
  {author} {\bibfnamefont {Z.}~\bibnamefont {Zhang}},\ and\ \bibinfo {author}
  {\bibfnamefont {Q.}~\bibnamefont {Zhuang}},\ }\bibfield  {title} {\bibinfo
  {title} {Entanglement-enhanced optomechanical sensor array with application
  to dark matter searches},\ }\bibfield  {journal} {\bibinfo  {journal}
  {Communications Physics}\ }\textbf {\bibinfo {volume} {6}},\ \href
  {https://doi.org/10.1038/s42005-023-01357-z} {10.1038/s42005-023-01357-z}
  (\bibinfo {year} {2023})\BibitemShut {NoStop}%
\bibitem [{\citenamefont {Djorwé}\ \emph
  {et~al.}(2024{\natexlab{a}})\citenamefont {Djorwé}, \citenamefont {Asjad},
  \citenamefont {Pennec}, \citenamefont {Dutykh},\ and\ \citenamefont
  {Djafari-Rouhani}}]{Djor2024}%
  \BibitemOpen
  \bibfield  {author} {\bibinfo {author} {\bibfnamefont {P.}~\bibnamefont
  {Djorwé}}, \bibinfo {author} {\bibfnamefont {M.}~\bibnamefont {Asjad}},
  \bibinfo {author} {\bibfnamefont {Y.}~\bibnamefont {Pennec}}, \bibinfo
  {author} {\bibfnamefont {D.}~\bibnamefont {Dutykh}},\ and\ \bibinfo {author}
  {\bibfnamefont {B.}~\bibnamefont {Djafari-Rouhani}},\ }\bibfield  {title}
  {\bibinfo {title} {Parametrically enhancing sensor sensitivity at an
  exceptional point},\ }\href
  {https://doi.org/10.1103/PhysRevResearch.6.033284} {\bibfield  {journal}
  {\bibinfo  {journal} {Physical Review Research}\ }\textbf {\bibinfo {volume}
  {6}},\ \bibinfo {pages} {033284} (\bibinfo {year}
  {2024}{\natexlab{a}})}\BibitemShut {NoStop}%
\bibitem [{\citenamefont {Barzanjeh}\ \emph {et~al.}(2021)\citenamefont
  {Barzanjeh}, \citenamefont {Xuereb}, \citenamefont {Gröblacher},
  \citenamefont {Paternostro}, \citenamefont {Regal},\ and\ \citenamefont
  {Weig}}]{Barzanjeh2021}%
  \BibitemOpen
  \bibfield  {author} {\bibinfo {author} {\bibfnamefont {S.}~\bibnamefont
  {Barzanjeh}}, \bibinfo {author} {\bibfnamefont {A.}~\bibnamefont {Xuereb}},
  \bibinfo {author} {\bibfnamefont {S.}~\bibnamefont {Gröblacher}}, \bibinfo
  {author} {\bibfnamefont {M.}~\bibnamefont {Paternostro}}, \bibinfo {author}
  {\bibfnamefont {C.~A.}\ \bibnamefont {Regal}},\ and\ \bibinfo {author}
  {\bibfnamefont {E.~M.}\ \bibnamefont {Weig}},\ }\bibfield  {title} {\bibinfo
  {title} {Optomechanics for quantum technologies},\ }\href
  {https://doi.org/10.1038/s41567-021-01402-0} {\bibfield  {journal} {\bibinfo
  {journal} {Nature Physics}\ }\textbf {\bibinfo {volume} {18}},\ \bibinfo
  {pages} {15} (\bibinfo {year} {2021})}\BibitemShut {NoStop}%
\bibitem [{\citenamefont {Roelli}\ \emph {et~al.}(2016)\citenamefont {Roelli},
  \citenamefont {Galland}, \citenamefont {Piro},\ and\ \citenamefont
  {Kippenberg}}]{Roelli2016}%
  \BibitemOpen
  \bibfield  {author} {\bibinfo {author} {\bibfnamefont {P.}~\bibnamefont
  {Roelli}}, \bibinfo {author} {\bibfnamefont {C.}~\bibnamefont {Galland}},
  \bibinfo {author} {\bibfnamefont {N.}~\bibnamefont {Piro}},\ and\ \bibinfo
  {author} {\bibfnamefont {T.~J.}\ \bibnamefont {Kippenberg}},\ }\bibfield
  {title} {\bibinfo {title} {Molecular cavity optomechanics as a theory of
  plasmon-enhanced raman scattering},\ }\href
  {https://doi.org/10.1038/nnano.2015.264} {\bibfield  {journal} {\bibinfo
  {journal} {Nature Nanotechnology}\ }\textbf {\bibinfo {volume} {11}},\
  \bibinfo {pages} {164} (\bibinfo {year} {2016})}\BibitemShut {NoStop}%
\bibitem [{\citenamefont {Esteban}\ \emph {et~al.}(2022)\citenamefont
  {Esteban}, \citenamefont {Baumberg},\ and\ \citenamefont
  {Aizpurua}}]{esteban2022molecular}%
  \BibitemOpen
  \bibfield  {author} {\bibinfo {author} {\bibfnamefont {R.}~\bibnamefont
  {Esteban}}, \bibinfo {author} {\bibfnamefont {J.~J.}\ \bibnamefont
  {Baumberg}},\ and\ \bibinfo {author} {\bibfnamefont {J.}~\bibnamefont
  {Aizpurua}},\ }\bibfield  {title} {\bibinfo {title} {Molecular optomechanics
  approach to surface-enhanced raman scattering},\ }\href
  {https://doi.org/10.1021/acs.accounts.1c00759} {\bibfield  {journal}
  {\bibinfo  {journal} {Accounts of Chemical Research}\ }\textbf {\bibinfo
  {volume} {55}},\ \bibinfo {pages} {1889} (\bibinfo {year}
  {2022})}\BibitemShut {NoStop}%
\bibitem [{\citenamefont {Schmidt}\ \emph {et~al.}(2017)\citenamefont
  {Schmidt}, \citenamefont {Esteban}, \citenamefont {Benz}, \citenamefont
  {Baumberg},\ and\ \citenamefont {Aizpurua}}]{schmidt2017}%
  \BibitemOpen
  \bibfield  {author} {\bibinfo {author} {\bibfnamefont {M.~K.}\ \bibnamefont
  {Schmidt}}, \bibinfo {author} {\bibfnamefont {R.}~\bibnamefont {Esteban}},
  \bibinfo {author} {\bibfnamefont {F.}~\bibnamefont {Benz}}, \bibinfo {author}
  {\bibfnamefont {J.~J.}\ \bibnamefont {Baumberg}},\ and\ \bibinfo {author}
  {\bibfnamefont {J.}~\bibnamefont {Aizpurua}},\ }\bibfield  {title} {\bibinfo
  {title} {Linking classical and molecular optomechanics descriptions of
  sers},\ }\href {https://doi.org/10.1039/c7fd00145b} {\bibfield  {journal}
  {\bibinfo  {journal} {Faraday Discussions}\ }\textbf {\bibinfo {volume}
  {205}},\ \bibinfo {pages} {31} (\bibinfo {year} {2017})}\BibitemShut
  {NoStop}%
\bibitem [{\citenamefont {Roelli}\ \emph {et~al.}(2024)\citenamefont {Roelli},
  \citenamefont {Hu}, \citenamefont {Verhagen}, \citenamefont {Reich},\ and\
  \citenamefont {Galland}}]{Roelli2024}%
  \BibitemOpen
  \bibfield  {author} {\bibinfo {author} {\bibfnamefont {P.}~\bibnamefont
  {Roelli}}, \bibinfo {author} {\bibfnamefont {H.}~\bibnamefont {Hu}}, \bibinfo
  {author} {\bibfnamefont {E.}~\bibnamefont {Verhagen}}, \bibinfo {author}
  {\bibfnamefont {S.}~\bibnamefont {Reich}},\ and\ \bibinfo {author}
  {\bibfnamefont {C.}~\bibnamefont {Galland}},\ }\bibfield  {title} {\bibinfo
  {title} {Nanocavities for molecular optomechanics: Their fundamental
  description and applications},\ }\href
  {https://doi.org/10.1021/acsphotonics.4c01548} {\bibfield  {journal}
  {\bibinfo  {journal} {ACS Photonics}\ }\textbf {\bibinfo {volume} {11}},\
  \bibinfo {pages} {4486} (\bibinfo {year} {2024})}\BibitemShut {NoStop}%
\bibitem [{\citenamefont {Chikkaraddy}\ \emph {et~al.}(2016)\citenamefont
  {Chikkaraddy}, \citenamefont {de~Nijs}, \citenamefont {Benz}, \citenamefont
  {Barrow}, \citenamefont {Scherman}, \citenamefont {Rosta}, \citenamefont
  {Demetriadou}, \citenamefont {Fox}, \citenamefont {Hess},\ and\ \citenamefont
  {Baumberg}}]{Chikkaraddy2016}%
  \BibitemOpen
  \bibfield  {author} {\bibinfo {author} {\bibfnamefont {R.}~\bibnamefont
  {Chikkaraddy}}, \bibinfo {author} {\bibfnamefont {B.}~\bibnamefont
  {de~Nijs}}, \bibinfo {author} {\bibfnamefont {F.}~\bibnamefont {Benz}},
  \bibinfo {author} {\bibfnamefont {S.~J.}\ \bibnamefont {Barrow}}, \bibinfo
  {author} {\bibfnamefont {O.~A.}\ \bibnamefont {Scherman}}, \bibinfo {author}
  {\bibfnamefont {E.}~\bibnamefont {Rosta}}, \bibinfo {author} {\bibfnamefont
  {A.}~\bibnamefont {Demetriadou}}, \bibinfo {author} {\bibfnamefont
  {P.}~\bibnamefont {Fox}}, \bibinfo {author} {\bibfnamefont {O.}~\bibnamefont
  {Hess}},\ and\ \bibinfo {author} {\bibfnamefont {J.~J.}\ \bibnamefont
  {Baumberg}},\ }\bibfield  {title} {\bibinfo {title} {Single-molecule strong
  coupling at room temperature in plasmonic nanocavities},\ }\href
  {https://doi.org/10.1038/nature17974} {\bibfield  {journal} {\bibinfo
  {journal} {Nature}\ }\textbf {\bibinfo {volume} {535}},\ \bibinfo {pages}
  {127} (\bibinfo {year} {2016})}\BibitemShut {NoStop}%
\bibitem [{\citenamefont {Carlon~Zambon}\ \emph {et~al.}(2022)\citenamefont
  {Carlon~Zambon}, \citenamefont {Denis}, \citenamefont {De~Oliveira},
  \citenamefont {Ravets}, \citenamefont {Ciuti}, \citenamefont {Favero},\ and\
  \citenamefont {Bloch}}]{CarlonZambon2022}%
  \BibitemOpen
  \bibfield  {author} {\bibinfo {author} {\bibfnamefont {N.}~\bibnamefont
  {Carlon~Zambon}}, \bibinfo {author} {\bibfnamefont {Z.}~\bibnamefont
  {Denis}}, \bibinfo {author} {\bibfnamefont {R.}~\bibnamefont {De~Oliveira}},
  \bibinfo {author} {\bibfnamefont {S.}~\bibnamefont {Ravets}}, \bibinfo
  {author} {\bibfnamefont {C.}~\bibnamefont {Ciuti}}, \bibinfo {author}
  {\bibfnamefont {I.}~\bibnamefont {Favero}},\ and\ \bibinfo {author}
  {\bibfnamefont {J.}~\bibnamefont {Bloch}},\ }\bibfield  {title} {\bibinfo
  {title} {Enhanced cavity optomechanics with quantum-well exciton
  polaritons},\ }\href {https://doi.org/10.1103/physrevlett.129.093603}
  {\bibfield  {journal} {\bibinfo  {journal} {Physical Review Letters}\
  }\textbf {\bibinfo {volume} {129}},\ \bibinfo {pages} {093603} (\bibinfo
  {year} {2022})}\BibitemShut {NoStop}%
\bibitem [{\citenamefont {Liu}\ \emph {et~al.}(2019)\citenamefont {Liu},
  \citenamefont {Zhao},\ and\ \citenamefont {Wu}}]{Liu2019}%
  \BibitemOpen
  \bibfield  {author} {\bibinfo {author} {\bibfnamefont {J.}~\bibnamefont
  {Liu}}, \bibinfo {author} {\bibfnamefont {Q.}~\bibnamefont {Zhao}},\ and\
  \bibinfo {author} {\bibfnamefont {N.}~\bibnamefont {Wu}},\ }\bibfield
  {title} {\bibinfo {title} {Vibration-assisted exciton transfer in molecular
  aggregates strongly coupled to confined light fields},\ }\bibfield  {journal}
  {\bibinfo  {journal} {The Journal of Chemical Physics}\ }\textbf {\bibinfo
  {volume} {150}},\ \href {https://doi.org/10.1063/1.5045706}
  {10.1063/1.5045706} (\bibinfo {year} {2019})\BibitemShut {NoStop}%
\bibitem [{\citenamefont {Huang}\ \emph {et~al.}(2024)\citenamefont {Huang},
  \citenamefont {Lei}, \citenamefont {Agarwal},\ and\ \citenamefont
  {Zhang}}]{Huang2024}%
  \BibitemOpen
  \bibfield  {author} {\bibinfo {author} {\bibfnamefont {J.}~\bibnamefont
  {Huang}}, \bibinfo {author} {\bibfnamefont {D.}~\bibnamefont {Lei}}, \bibinfo
  {author} {\bibfnamefont {G.~S.}\ \bibnamefont {Agarwal}},\ and\ \bibinfo
  {author} {\bibfnamefont {Z.}~\bibnamefont {Zhang}},\ }\bibfield  {title}
  {\bibinfo {title} {Collective quantum entanglement in molecular cavity
  optomechanics},\ }\href {https://doi.org/10.1103/PhysRevB.110.184306}
  {\bibfield  {journal} {\bibinfo  {journal} {Phys. Rev. B}\ }\textbf {\bibinfo
  {volume} {110}},\ \bibinfo {pages} {184306} (\bibinfo {year}
  {2024})}\BibitemShut {NoStop}%
\bibitem [{\citenamefont {Xiang}\ and\ \citenamefont
  {Xiong}(2024)}]{Xiang2024}%
  \BibitemOpen
  \bibfield  {author} {\bibinfo {author} {\bibfnamefont {B.}~\bibnamefont
  {Xiang}}\ and\ \bibinfo {author} {\bibfnamefont {W.}~\bibnamefont {Xiong}},\
  }\bibfield  {title} {\bibinfo {title} {Molecular polaritons for chemistry,
  photonics and quantum technologies},\ }\href
  {https://doi.org/10.1021/acs.chemrev.3c00662} {\bibfield  {journal} {\bibinfo
   {journal} {Chemical Reviews}\ }\textbf {\bibinfo {volume} {124}},\ \bibinfo
  {pages} {2512} (\bibinfo {year} {2024})}\BibitemShut {NoStop}%
\bibitem [{\citenamefont {Emale}\ \emph {et~al.}(2025)\citenamefont {Emale},
  \citenamefont {Peng}, \citenamefont {Djorwé}, \citenamefont {Sarma},
  \citenamefont {Abdourahimi}, \citenamefont {Abdel-Aty}, \citenamefont
  {Nisar},\ and\ \citenamefont {Engo}}]{EMALE2025416919}%
  \BibitemOpen
  \bibfield  {author} {\bibinfo {author} {\bibfnamefont {K.}~\bibnamefont
  {Emale}}, \bibinfo {author} {\bibfnamefont {J.-X.}\ \bibnamefont {Peng}},
  \bibinfo {author} {\bibfnamefont {P.}~\bibnamefont {Djorwé}}, \bibinfo
  {author} {\bibfnamefont {A.}~\bibnamefont {Sarma}}, \bibinfo {author}
  {\bibnamefont {Abdourahimi}}, \bibinfo {author} {\bibfnamefont {A.-H.}\
  \bibnamefont {Abdel-Aty}}, \bibinfo {author} {\bibfnamefont {K.}~\bibnamefont
  {Nisar}},\ and\ \bibinfo {author} {\bibfnamefont {S.}~\bibnamefont {Engo}},\
  }\bibfield  {title} {\bibinfo {title} {Quantum correlations enhanced in
  hybrid optomechanical system via phase tuning},\ }\href
  {https://doi.org/https://doi.org/10.1016/j.physb.2025.416919} {\bibfield
  {journal} {\bibinfo  {journal} {Physica B: Condensed Matter}\ }\textbf
  {\bibinfo {volume} {701}},\ \bibinfo {pages} {416919} (\bibinfo {year}
  {2025})}\BibitemShut {NoStop}%
\bibitem [{\citenamefont {Massembele}\ \emph {et~al.}(2024)\citenamefont
  {Massembele}, \citenamefont {Djorw\'e}, \citenamefont {Sarma}, \citenamefont
  {Abdel-Aty},\ and\ \citenamefont {Engo}}]{Massembele2024}%
  \BibitemOpen
  \bibfield  {author} {\bibinfo {author} {\bibfnamefont {D.~R.~K.}\
  \bibnamefont {Massembele}}, \bibinfo {author} {\bibfnamefont
  {P.}~\bibnamefont {Djorw\'e}}, \bibinfo {author} {\bibfnamefont {A.~K.}\
  \bibnamefont {Sarma}}, \bibinfo {author} {\bibfnamefont {A.-H.}\ \bibnamefont
  {Abdel-Aty}},\ and\ \bibinfo {author} {\bibfnamefont {S.~G.~N.}\ \bibnamefont
  {Engo}},\ }\bibfield  {title} {\bibinfo {title} {Quantum entanglement
  assisted via duffing nonlinearity},\ }\href
  {https://doi.org/10.1103/PhysRevA.110.043502} {\bibfield  {journal} {\bibinfo
   {journal} {Phys. Rev. A}\ }\textbf {\bibinfo {volume} {110}},\ \bibinfo
  {pages} {043502} (\bibinfo {year} {2024})}\BibitemShut {NoStop}%
\bibitem [{\citenamefont {Djorwé}\ \emph
  {et~al.}(2024{\natexlab{b}})\citenamefont {Djorwé}, \citenamefont
  {Abdel-Aty}, \citenamefont {Nisar},\ and\ \citenamefont {Engo}}]{Djo2024}%
  \BibitemOpen
  \bibfield  {author} {\bibinfo {author} {\bibfnamefont {P.}~\bibnamefont
  {Djorwé}}, \bibinfo {author} {\bibfnamefont {A.-H.}\ \bibnamefont
  {Abdel-Aty}}, \bibinfo {author} {\bibfnamefont {K.}~\bibnamefont {Nisar}},\
  and\ \bibinfo {author} {\bibfnamefont {S.}~\bibnamefont {Engo}},\ }\bibfield
  {title} {\bibinfo {title} {Optomechanical entanglement induced by backward
  stimulated brillouin scattering},\ }\href
  {https://doi.org/10.1016/j.ijleo.2024.172097} {\bibfield  {journal} {\bibinfo
   {journal} {Optik}\ }\textbf {\bibinfo {volume} {319}},\ \bibinfo {pages}
  {172097} (\bibinfo {year} {2024}{\natexlab{b}})}\BibitemShut {NoStop}%
\bibitem [{\citenamefont {Zhang}\ \emph {et~al.}(2018)\citenamefont {Zhang},
  \citenamefont {Liu}, \citenamefont {Yang},\ and\ \citenamefont
  {Zhang}}]{Zhang2018}%
  \BibitemOpen
  \bibfield  {author} {\bibinfo {author} {\bibfnamefont {J.}~\bibnamefont
  {Zhang}}, \bibinfo {author} {\bibfnamefont {X.}~\bibnamefont {Liu}}, \bibinfo
  {author} {\bibfnamefont {R.}~\bibnamefont {Yang}},\ and\ \bibinfo {author}
  {\bibfnamefont {T.}~\bibnamefont {Zhang}},\ }\bibfield  {title} {\bibinfo
  {title} {Scheme for enhancing quadripartite entangled optical modes from an
  opto-mechanical system},\ }\href {https://doi.org/10.1364/JOSAB.35.002945}
  {\bibfield  {journal} {\bibinfo  {journal} {J. Opt. Soc. Am. B}\ }\textbf
  {\bibinfo {volume} {35}},\ \bibinfo {pages} {2945} (\bibinfo {year}
  {2018})}\BibitemShut {NoStop}%
\bibitem [{\citenamefont {Pan}\ \emph {et~al.}(2022)\citenamefont {Pan},
  \citenamefont {Xiao},\ and\ \citenamefont {Zhai}}]{Pan2022}%
  \BibitemOpen
  \bibfield  {author} {\bibinfo {author} {\bibfnamefont {G.}~\bibnamefont
  {Pan}}, \bibinfo {author} {\bibfnamefont {R.}~\bibnamefont {Xiao}},\ and\
  \bibinfo {author} {\bibfnamefont {C.}~\bibnamefont {Zhai}},\ }\bibfield
  {title} {\bibinfo {title} {Entanglement and output squeezing in a distant
  nano-electro-optomechanical system generated by optical parametric
  amplifiers},\ }\href {https://doi.org/10.1088/1612-202X/ac5e3a} {\bibfield
  {journal} {\bibinfo  {journal} {Laser Physics Letters}\ }\textbf {\bibinfo
  {volume} {19}},\ \bibinfo {pages} {055203} (\bibinfo {year}
  {2022})}\BibitemShut {NoStop}%
\bibitem [{\citenamefont {Kibret}\ \emph {et~al.}(2023)\citenamefont {Kibret},
  \citenamefont {Derge},\ and\ \citenamefont {Tesfahannes}}]{Kibret2023}%
  \BibitemOpen
  \bibfield  {author} {\bibinfo {author} {\bibfnamefont {A.~A.}\ \bibnamefont
  {Kibret}}, \bibinfo {author} {\bibfnamefont {T.~Y.}\ \bibnamefont {Derge}},\
  and\ \bibinfo {author} {\bibfnamefont {T.~G.}\ \bibnamefont {Tesfahannes}},\
  }\bibfield  {title} {\bibinfo {title} {Steady-state entanglement in a hybrid
  optomechanical system enhanced by optical parametric amplifiers},\ }\href
  {https://doi.org/10.1364/OPTCON.502349} {\bibfield  {journal} {\bibinfo
  {journal} {Opt. Continuum}\ }\textbf {\bibinfo {volume} {2}},\ \bibinfo
  {pages} {2131} (\bibinfo {year} {2023})}\BibitemShut {NoStop}%
\bibitem [{\citenamefont {Chitsazi}\ \emph {et~al.}(2017)\citenamefont
  {Chitsazi}, \citenamefont {Li}, \citenamefont {Ellis},\ and\ \citenamefont
  {Kottos}}]{Chitsazi2017}%
  \BibitemOpen
  \bibfield  {author} {\bibinfo {author} {\bibfnamefont {M.}~\bibnamefont
  {Chitsazi}}, \bibinfo {author} {\bibfnamefont {H.}~\bibnamefont {Li}},
  \bibinfo {author} {\bibfnamefont {F.}~\bibnamefont {Ellis}},\ and\ \bibinfo
  {author} {\bibfnamefont {T.}~\bibnamefont {Kottos}},\ }\bibfield  {title}
  {\bibinfo {title} {Experimental realization of floquet pt -symmetric
  systems},\ }\href {https://doi.org/10.1103/physrevlett.119.093901} {\bibfield
   {journal} {\bibinfo  {journal} {Physical Review Letters}\ }\textbf {\bibinfo
  {volume} {119}},\ \bibinfo {pages} {093901} (\bibinfo {year}
  {2017})}\BibitemShut {NoStop}%
\bibitem [{\citenamefont {Gardiner}\ and\ \citenamefont
  {Zoller}(2004)}]{Gardiner2004}%
  \BibitemOpen
  \bibfield  {author} {\bibinfo {author} {\bibfnamefont {C.}~\bibnamefont
  {Gardiner}}\ and\ \bibinfo {author} {\bibfnamefont {P.}~\bibnamefont
  {Zoller}},\ }\href {https://doi.org/10.1142/9781783264629} {\emph {\bibinfo
  {title} {Quantum noise: A Handbook of Markovian and Non-Markovian Quantum
  Stochastic Methods with Applications to Quantum Optics}}},\ \bibinfo
  {edition} {3rd}\ ed.,\ Springer series in synergetics\ (\bibinfo  {publisher}
  {Springer},\ \bibinfo {year} {2004})\ pp.\ \bibinfo {pages}
  {1--311}\BibitemShut {NoStop}%
\bibitem [{\citenamefont {DeJesus}\ and\ \citenamefont
  {Kaufman}(1987)}]{DeJesus1987}%
  \BibitemOpen
  \bibfield  {author} {\bibinfo {author} {\bibfnamefont {E.~X.}\ \bibnamefont
  {DeJesus}}\ and\ \bibinfo {author} {\bibfnamefont {C.}~\bibnamefont
  {Kaufman}},\ }\bibfield  {title} {\bibinfo {title} {Routh-hurwitz criterion
  in the examination of eigenvalues of a system of nonlinear ordinary
  differential equations},\ }\href {https://doi.org/10.1103/PhysRevA.35.5288}
  {\bibfield  {journal} {\bibinfo  {journal} {Phys. Rev. A}\ }\textbf {\bibinfo
  {volume} {35}},\ \bibinfo {pages} {5288} (\bibinfo {year}
  {1987})}\BibitemShut {NoStop}%
\bibitem [{\citenamefont {Plenio}(2005)}]{Plenio2005}%
  \BibitemOpen
  \bibfield  {author} {\bibinfo {author} {\bibfnamefont {M.~B.}\ \bibnamefont
  {Plenio}},\ }\bibfield  {title} {\bibinfo {title} {Logarithmic negativity: A
  full entanglement monotone that is not convex},\ }\href
  {https://doi.org/10.1103/PhysRevLett.95.090503} {\bibfield  {journal}
  {\bibinfo  {journal} {Phys. Rev. Lett.}\ }\textbf {\bibinfo {volume} {95}},\
  \bibinfo {pages} {090503} (\bibinfo {year} {2005})}\BibitemShut {NoStop}%
\bibitem [{\citenamefont {Xiao}\ \emph {et~al.}(2012)\citenamefont {Xiao},
  \citenamefont {Liu}, \citenamefont {Li}, \citenamefont {Chen}, \citenamefont
  {Li},\ and\ \citenamefont {Gong}}]{Xiao2012}%
  \BibitemOpen
  \bibfield  {author} {\bibinfo {author} {\bibfnamefont {Y.-F.}\ \bibnamefont
  {Xiao}}, \bibinfo {author} {\bibfnamefont {Y.-C.}\ \bibnamefont {Liu}},
  \bibinfo {author} {\bibfnamefont {B.-B.}\ \bibnamefont {Li}}, \bibinfo
  {author} {\bibfnamefont {Y.-L.}\ \bibnamefont {Chen}}, \bibinfo {author}
  {\bibfnamefont {Y.}~\bibnamefont {Li}},\ and\ \bibinfo {author}
  {\bibfnamefont {Q.}~\bibnamefont {Gong}},\ }\bibfield  {title} {\bibinfo
  {title} {Strongly enhanced light-matter interaction in a hybrid
  photonic-plasmonic resonator},\ }\href
  {https://doi.org/10.1103/PhysRevA.85.031805} {\bibfield  {journal} {\bibinfo
  {journal} {Phys. Rev. A}\ }\textbf {\bibinfo {volume} {85}},\ \bibinfo
  {pages} {031805} (\bibinfo {year} {2012})}\BibitemShut {NoStop}%
\bibitem [{\citenamefont {Liu}\ and\ \citenamefont {Zhu}(2017)}]{Liu2017}%
  \BibitemOpen
  \bibfield  {author} {\bibinfo {author} {\bibfnamefont {J.}~\bibnamefont
  {Liu}}\ and\ \bibinfo {author} {\bibfnamefont {K.-D.}\ \bibnamefont {Zhu}},\
  }\bibfield  {title} {\bibinfo {title} {Coupled quantum molecular cavity
  optomechanics with surface plasmon enhancement},\ }\href
  {https://doi.org/10.1364/PRJ.5.000450} {\bibfield  {journal} {\bibinfo
  {journal} {Photon. Res.}\ }\textbf {\bibinfo {volume} {5}},\ \bibinfo {pages}
  {450} (\bibinfo {year} {2017})}\BibitemShut {NoStop}%
\bibitem [{\citenamefont {Zou}\ \emph {et~al.}(2024)\citenamefont {Zou},
  \citenamefont {Du}, \citenamefont {Li},\ and\ \citenamefont
  {Dong}}]{Zou2024}%
  \BibitemOpen
  \bibfield  {author} {\bibinfo {author} {\bibfnamefont {F.}~\bibnamefont
  {Zou}}, \bibinfo {author} {\bibfnamefont {L.}~\bibnamefont {Du}}, \bibinfo
  {author} {\bibfnamefont {Y.}~\bibnamefont {Li}},\ and\ \bibinfo {author}
  {\bibfnamefont {H.}~\bibnamefont {Dong}},\ }\bibfield  {title} {\bibinfo
  {title} {Amplifying frequency up-converted infrared signals with a molecular
  optomechanical cavity},\ }\href
  {https://doi.org/10.1103/PhysRevLett.132.153602} {\bibfield  {journal}
  {\bibinfo  {journal} {Phys. Rev. Lett.}\ }\textbf {\bibinfo {volume} {132}},\
  \bibinfo {pages} {153602} (\bibinfo {year} {2024})}\BibitemShut {NoStop}%
\bibitem [{\citenamefont {Schmidt}\ and\ \citenamefont
  {Steel}(2024)}]{Schmidt2024}%
  \BibitemOpen
  \bibfield  {author} {\bibinfo {author} {\bibfnamefont {M.~K.}\ \bibnamefont
  {Schmidt}}\ and\ \bibinfo {author} {\bibfnamefont {M.~J.}\ \bibnamefont
  {Steel}},\ }\bibfield  {title} {\bibinfo {title} {Molecular optomechanics in
  the anharmonic regime: from nonclassical mechanical states to mechanical
  lasing},\ }\href {https://doi.org/10.1088/1367-2630/ad32e4} {\bibfield
  {journal} {\bibinfo  {journal} {New Journal of Physics}\ }\textbf {\bibinfo
  {volume} {26}},\ \bibinfo {pages} {033041} (\bibinfo {year}
  {2024})}\BibitemShut {NoStop}%
\bibitem [{\citenamefont {Djorwe}\ \emph {et~al.}(2022)\citenamefont {Djorwe},
  \citenamefont {Yves Effa},\ and\ \citenamefont
  {G. Nana Engo}}]{Djorwe2022}%
  \BibitemOpen
  \bibfield  {author} {\bibinfo {author} {\bibfnamefont {P.}~\bibnamefont
  {Djorwe}}, \bibinfo {author} {\bibfnamefont {J.}~\bibnamefont {Yves Effa}},\
  and\ \bibinfo {author} {\bibfnamefont {S.}~\bibnamefont {G. Nana Engo}},\
  }\bibfield  {title} {\bibinfo {title} {Hidden attractors and metamorphoses of
  basin boundaries in optomechanics},\ }\href
  {https://doi.org/10.1007/s11071-022-08139-2} {\bibfield  {journal} {\bibinfo
  {journal} {Nonlinear Dynamics}\ }\textbf {\bibinfo {volume} {111}},\ \bibinfo
  {pages} {5905} (\bibinfo {year} {2022})}\BibitemShut {NoStop}%
\end{thebibliography}%
\end{document}